 \newcommand{\articlereport}[2]{#2} % comment this for article
\articlereport
{
\documentclass{llncs}
}
{
\documentclass{article}
}
\articlereport{}{\usepackage[left=2cm,top=2cm,right=3cm]{geometry}} % RT
\usepackage{fancybox}
\usepackage{amssymb}
\usepackage{amsfonts}
\usepackage{amsmath}
\articlereport{}{\usepackage[amsmath,thmmarks]{ntheorem}}
\usepackage{pstricks}
\usepackage{ifthen}
\usepackage{scalefnt}
\usepackage{url}
\usepackage{verbatim}
\usepackage[small,compact]{titlesec}
\articlereport{}{
\newtheorem{theorem}{Theorem}%[section]
\newtheorem{lemma}[theorem]{Lemma}
\newtheorem{corollary}[theorem]{Corollary}
\newtheorem{proposition}[theorem]{Proposition}
{\theorembodyfont{\upshape}
  \theoremsymbol{\ensuremath{\Diamond}}
  \newtheorem{definition}[theorem]{Definition}
  
  \theoremsymbol{\ensuremath{\clubsuit}}
  \newtheorem{example}[theorem]{Example}
  \theoremstyle{nonumberplain}
}
\theoremheaderfont{\sc}\theorembodyfont{\upshape}
\theoremstyle{nonumberplain}
\theoremsymbol{\rule{1ex}{1ex}}
\newtheorem{proof}{Proof.}
}

\newcommand{\set}[1]{\{#1\}}
\newcommand{\tuple}[1]{\langle #1 \rangle}

\newcommand{\setof}[2]{\{#1\,|\:#2\}}

\newcommand{\dN}{\mathbb{N}}
\newcommand{\dZ}{\mathbb{Z}}
\newcommand{\myref}[1]{(\ref{#1})}

\newcommand{\sbtat}[2]{#1|_{#2}}
\newcommand{\iseq}{\simeq} % syntactic eq
\newcommand{\niseq}{\not \iseq}
\newcommand{\var}{\mbox{\it V}} % variables
\newcommand{\Var}[1]{\mbox{\it V}(#1)} % variables in Mnacho version
\newcommand{\someq}{\bowtie} % generic notation for = or !=

\newcommand{\select}{\texttt{select}}
\newcommand{\su}{\mathrm{s}}

\newcommand{\store}{\texttt{store}}

\newcommand{\thearr}{\mathcal{A}}
\newcommand{\thearrz}{{\thearr_{\dZ}}}

\newcommand{\prc}{\mathrm{p}}

\newcommand{\speconst}{\chi}
\newcommand{\abscl}[3]{#1\,\|\, #2\rightarrow #3}
\newcommand{\absclause}{\abscl{\Lambda}{\Gamma}{\Delta}}
\newcommand{\absclempty}[1]{\abscl{#1}{}{}}
\newcommand{\spsub}{\unlhd}
\newcommand{\absvar}[1]{\var_{\mathrm{abs}}(#1)}
\newcommand{\ineqvar}[1]{\var_{\mathrm{ineq}}(#1)}
\newcommand{\preconstrained}{preconstrained}
\newcommand{\absterm}[1]{\mathrm{T}_{\mathrm{abs}}(#1)}
\newcommand{\bdef}{$B$-definition}
\newcommand{\setbdefs}[2]{\Theta_{#1}[#2]}
\newcommand{\addabs}[2]{[#1,#2]}
\newcommand{\addsetabs}[2]{#1_{#2}}
\newcommand{\irule}[2]{\begin{array}{c}{#1}\\\hline{#2}\end{array}}
\newcommand{\srule}[2]{\begin{array}{c}{#1}\\\hline\hline{#2}\end{array}}

\newcommand{\emptcl}{\square}

\newcommand{\geninst}[2]{#1\set{#2}}
\newcommand{\lfta}{\leftarrow}
\newcommand{\replat}[3]{#1[#2]_{#3}}
\newcommand{\mapspos}[2]{\tuple{#1,#2}}

\newcommand{\modelz}{\models_{\dZ}}
\newcommand{\isless}{\preceq}
\newcommand{\mybsup}[3]{\bar{#1}_{#2}^{#3}}
\newcommand{\sorts}{{\mathcal S}}
\newcommand{\asort}{{\tt s}}
\newcommand{\upbnd}[2]{B_{#1}^{#2}}
\newcommand{\myabscl}[2]{#1\,\|\, #2}
\newcommand{\intt}{\Theta}

\newcommand{\itt}[1]{T_{\dZ}(#1)}
\newcommand{\aclass}{{\mathcal C}}
\newcommand{\instgamma}{\gamma}

\newcommand{\instcl}[2]{\instgamma_{#1}^{#2}} % Niko: need two arguments for this
\newcommand{\disc}{independent}
\newcommand{\infsys}{\mathcal{H}}
\newcommand{\sttwo}{\text{\it St}_2}
\newcommand{\stzero}{\text{\it St}_0}

\newcommand{\lev}{\mathrm{level}}
\newcommand{\im}[1]{\mbox{Im}[#1]}
\newcommand{\tr}[1]{#1\!\!\downarrow_{0}}
\newcommand{\normI}{\!\!\downarrow_I}
\newcommand{\spinc}{\sqsubseteq_{\dZ}}

\begin{document}
\articlereport
{
\title{Instantiation of SMT problems modulo Integers}
\author{Mnacho Echenim \and Nicolas Peltier}
\institute{University of Grenoble\thanks{emails: \texttt{Mnacho.Echenim@imag.fr} and \texttt{Nicolas.Peltier@imag.fr}} (LIG, Grenoble INP/CNRS)}
}
{
\title{Instantiation of SMT problems modulo Integers\\ (technical report)}
\author{Mnacho Echenim \and Nicolas Peltier}
\date{March 2010}
 }

\articlereport{\scalefont{0.971}}{}

\maketitle

\begin{abstract}
  Many decision procedures for SMT problems rely more or less
  implicitly on an instantiation of the axioms of the theories under
  consideration, and differ by making use of the additional properties of
  each theory, in order to increase efficiency. We present a new
  technique for devising complete instantiation schemes on SMT
  problems over a combination of linear arithmetic with another
  theory ${\mathcal T}$. The method consists in first instantiating the arithmetic
  part of the formula, and then getting rid of the remaining variables in
  the problem by using an instantiation strategy which is complete for
  ${\mathcal T}$.
  We provide examples evidencing that not only is this technique
  generic (in the sense that it applies to a wide range of theories)
  but it is also efficient, even compared to state-of-the-art
  instantiation schemes for specific theories.
\end{abstract}

%\begin{keywords}
%Automated reasoning, linear arithmetic, satisfiability modulo theories
%(SMT), instantiation-based proof pro\-ce\-dures,
% theory of arrays,  stratified clause sets.
%\end{keywords}

\articlereport{}{\section{introduction}
}
Research in the domain of Satisfiability Modulo Theories focuses on
the design of decision procedures capable of testing the
satisfiability of ground formulas modulo a given background
theory. Such satisfiability checks may arise as a subproblem during
the task of proving a more general formula in, e.g., software
verification or interactive theorem proving. The background theories
under consideration may define usual mathematical objects such as
linear arithmetic, or data structures such as arrays or lists. The
tools that implement these decision procedures are named SMT solvers,
and they are designed to be as efficient as possible. This efficiency
is obtained thanks to a sophisticated combination of state-of-the-art
techniques derived from SAT solving, and ad-hoc procedures designed to
handle each specific theory (see, e.g., \cite{BSST09} for a survey).

The lack of genericity of these theory solvers may become an issue, as
additional theories, either new ones or extensions of former ones, are
defined. For instance, a programmer may wish to add new axioms to the
usual theory of arrays to specify, e.g., dimensions, sortedness, or
definition domains. A solution to this lack of genericity was
investigated in \cite{ARR03,ABRS09}, where a first-order theorem
prover is used to solve SMT problems. Once it is proved that the
theorem prover terminates on SMT problems for a given theory, it can
be used as an SMT solver for that theory, and no additional
implementation is required. Also, under certain conditions such as
variable-inactivity (see, e.g., \cite{ABRS09,viPolyTsat}), the
theorem prover can also be used as an SMT prover for a combination of
theories at no further expense. However, first-order theorem provers
are not capable of efficiently handling the potentially large boolean
structures of SMT problems. A solution to this problem was proposed in
\cite{dpByStages}, with an approach consisting of decomposing an SMT
problem in such a way that the theorem prover does not need to handle
its boolean part. But even with this approach, theorem provers do not
seem capable to compete with state-of-the-art SMT solvers.

A new approach to handling the genericity issue consists in devising a
general instantiation scheme for SMT problems. The principle of this
approach is to instantiate the axioms of the theories so that it is
only necessary to feed a \emph{ground} formula to the SMT solver. The
problem is then to find a way to instantiate the axioms as little as
possible so that the size of the resulting formula does not blow up,
and still retain completeness: the instantiated set of clauses must be
satisfiable if and only if the original
set is. Such an approach was
investigated in \cite{EP09}, and an instantiation scheme was devised
along with a syntactic characterization of theories for which it is
refutationally complete.
One theory that cannot be handled by this approach is the theory of
\emph{linear arithmetic},
which is infinitely axiomatized. Yet, this theory
frequently appears in SMT problems, such as the problems on arrays
with integer indices. Handling linear arithmetic is also a challenge
in first-order theorem proving, and several systems have been designed
to handle the arithmetic parts of the formulas in an efficient way
(see, e.g., \cite{Korovin07} or the calculus of \cite{AlthausKW09},
which derives from \cite{BachmairGW94}).

In this paper, we
devise
an
instantiation scheme for theories containing particular integer
constraints. This scheme, together with that of \cite{EP09}, permits
to test the satisfiability of an SMT problem over a combination of
linear arithmetic with another theory, by feeding a ground formula to
an SMT solver. We show the potential efficiency of this scheme by
applying it to problems in the theory of \emph{arrays with integer indices},
and we show that it can generate sets of ground formulas that are much
smaller than the ones generated by the instantiation rule of
\cite{BradleyBook}.  To emphasize the genericity of our approach,
we also use it to integrate
  arithmetic constraints into a decidable subclass of many-sorted
  logic.

The paper is organized as follows. After recalling basic definitions
from automated theorem proving, we introduce the notion of
$\dZ$-clauses, which are a restriction of the abstracted
clauses of \cite{BachmairGW94,AlthausKW09}, along with the inference system
introduced in \cite{AlthausKW09}. We define a way of instantiating
integer variables in particular formulas, and show how to determine a
set of terms large enough to ensure completeness of the instantiation
technique on an SMT problem. We then prove that under some conditions
which are fulfilled by the scheme of \cite{EP09}, completeness is
retained after using the scheme to instantiate the remaining variables in the SMT
problems.
We conclude by showing how this combined scheme can be
applied on concrete problems.
\articlereport{Due to a lack of space, we did not
include the proofs in this paper;
the detailed proofs can all be found in a technical report available at \url{http://membres-lig.imag.fr/peltier/inst_la.pdf}.
The information on
the instantiation method for non integer variables is available in \cite{EPTR09}.
}{}

\articlereport
{
\newcommand{\sv}[1]{}
}
{
\newcommand{\sv}[1]{#1}
}
\section{Preliminaries}

We employ a many-sorted framework.  Let $\sorts$ denote a set of sorts,
containing in particular a symbol $\dZ$ denoting integers.  Every
variable is mapped to a unique sort in $\sorts$ and every function
symbol $f$ is mapped to a unique profile of the form $\asort_1 \times
\ldots \times \asort_n \rightarrow \asort$, where
$\asort_1,\ldots,\asort_n,\asort \in \sorts$ (possibly with $n = 0$);
the sort $\asort$ is the \emph{range} of the function $f$.
Terms are built with the usual conditions of well-sortedness.
The
signature contains in particular the symbols $0,-,+$ of respective profiles
$\rightarrow \dZ$, $\dZ \rightarrow \dZ$, $\dZ \times \dZ \rightarrow
\dZ$. The terms $s^i(0)$, $t+s(0)$, $t+(-s(0))$ and
$t+(-s)$ are abbreviated
by $i,\su(t),\prc(t)$ and $t-s$
respectively.
Terms (resp. variables) of sort $\dZ$ are called \emph{integer terms}
(resp. \emph{integer variables}).
A term is \emph{ground} if it contains no variable.  We assume that
there exists at least one ground term of each sort
and that for every function symbol of profile $\asort_1 \times \ldots
\times \asort_n \rightarrow \dZ$, we have $\asort_i = \dZ$ for all $i
\in [1..n]$: integer terms may only have integer subterms. \label{dep}
\sv{
%This implies that all the subterms of a term of sort
%$\dZ$ are integer terms.
In other words, a noninteger term may depend
on an integer term, but integer terms depend only on integer
terms. This condition imposes some sort of hierarchical stratification
between the theory of linear arithmetic and the other theory in which
the problem is solved.  As we shall see, this stratification plays a
crucial role in our approach.}

An \emph{atom} is either of the form $t
\isless s$ where $t,s$ are two terms of sort $\dZ$, or of the form $t
\iseq s$ where $t,s$ are terms of the same sort. An atom\footnote{The symbol $\someq$ represents either $\iseq$ or $\isless$.} $t \someq s$
is \emph{arithmetic} if $t,s$ are of sort
$\dZ$.  A \emph{clause} is an expression
of the form $\Gamma \rightarrow \Delta$, where $\Gamma,\Delta$ are
sequences of non-arithmetic atoms. A \emph{substitution} $\sigma$ is a function mapping every variable
$x$ to a term $x\sigma$ of the same sort. Substitution $\sigma$ is
\emph{ground} if for every variable $x$ in the domain of $\sigma$,
$x\sigma$ is ground. For any expression $\mathcal{E}$ (term, atom,  sequence of
atoms or clause), $\Var{\mathcal{E}}$ is the set of variables
occurring in $\mathcal{E}$ and $\mathcal{E}\sigma$ denotes the expression obtained by replacing
in $\mathcal{E}$ every variable $x$ in the domain of $\sigma$ by the term
$x\sigma$.
Interpretations are defined as usual. A \emph{$\dZ$-interpretation}
$I$ is an interpretation such that the domain of sort $\dZ$ is the set
of integers, and that the interpretation of the symbols $0,-,+$ is
defined as follows: $I(0) = 0, I(t+s) = I(t)+I(s)$ and $I(-t) =
-I(t)$.
 A ground atom $A$ is
\emph{satisfied} by an interpretation $I$ if either $A$ is of the form
$t \isless s$ and $I(t) \leq I(s)$ or $A$ is of the form $t \iseq s$
and $I(t) = I(s)$.  A clause $\Gamma \rightarrow \Delta$ is
\emph{satisfied} by an interpretation $I$ if for every ground
substitution $\sigma$, either there exists an atom $A \in
\Gamma\sigma$ that is \textbf{not} satisfied by $I$, or there exists an
atom $A \in \Delta\sigma$ that is satisfied by $I$.  A set of clauses
$S$ is \emph{satisfied} by $I$ if $I$ satisfies every clause in $S$.
As usual, we write $I \models S$ if $S$ is satisfied by $I$ and $S_1
\models S_2$ if every interpretation that satisfies $S_1$ also
satisfies $S_2$.
$S_1$ and $S_2$ are \emph{equivalent} if $S_1 \models S_2$ and $S_2
\models S_1$. We note $I \modelz S$ if $I$ is a $\dZ$-interpretation
that satisfies $S$; $S_1\modelz S_2$ if every $\dZ$-interpretation
satisfying $S_1$ also satisfies $S_2$, and $S_1, S_2$ are
\emph{$\dZ$-equivalent} if $S_1\modelz S_2$ and $S_2\modelz S_1$.

We assume the standard notions of positions in terms, atoms and
clauses.
As usual, given two terms $t$ and $s$, $\sbtat{t}{p}$ is the subterm occurring at position $p$ in $t$ and $\replat{t}{s}{p}$
denotes the term obtained from $t$ by replacing the subterm at
position $p$ by $s$.
Given an expression $\mathcal{E}$ (term, atom, clause...), a position
$p$ is a \emph{variable position in $\mathcal{E}$} if $\sbtat{\mathcal{E}}{p}$ is a
variable.

\begin{comment}
\begin{definition}
  Given a  term $t$, let $\set{p_1, \ldots,
    p_n}$ denote a set of variable positions in $t$. We
  define
  $\geninst{t}{\mapspos{p_1}{s_1}, \ldots, \mapspos{p_n}{s_n}}\ =\
    t[s_1]_{p_1}[s_2]_{p_2}\cdots[s_n]_{p_n}$.
  This notation is extended to atoms and clauses in the standard way.
\end{definition}

Note that $\geninst{t}{\mapspos{p_1}{s_1}, \ldots,
  \mapspos{p_n}{s_n}}$ is well-defined since for no two positions in
$\set{p_1,\ldots,p_n}$ is one the strict prefix of the other.

\begin{example}
  Let $t = f(g(x,y), h(x,z), y)$. The set of variable positions in $t$
  is $\set{1.1, 1.2, 2.1, 2.2, 3}$, and $\geninst{t}{\mapspos{1.1}{a},
  \mapspos{2.1}{b}, \mapspos{3}{c}} = f(g(a,y), h(b,z), c)$.
\end{example}
\end{comment}

The \emph{flattening} operation on a set of clauses $S$ consists in
replacing non constant ground terms $t$ occurring in $S$ by  fresh
constants $c$, and adding to $S$ the unit clause $t \iseq c$. We refer
the reader to,
e.g., \cite{ARR03} for more details.

\section{$\dZ$-clauses}

We introduce the class of $\dZ$-clauses. These are restricted
versions of the abstracted clauses of \cite{BachmairGW94,AlthausKW09},
as we impose that the arithmetic constraints be represented by atoms,
and not literals. We add
 this restriction for the sake of
readability; in fact it incurs no loss of generality:
for example, a
literal $\neg (a \isless b)$ can be replaced by the $\dZ$-equivalent
arithmetic atom $b \isless \prc(a)$.
We present some terminology from \cite{AlthausKW09}, adapted to our setting.

\begin{definition}
A \emph{$\dZ$-clause} is an expression of the form $\absclause$, where:
\begin{itemize}
\item{$\Lambda$ is a sequence of arithmetic atoms (the
    \emph{arithmetic part} of $\absclause$);}
\item{$\Gamma\rightarrow\Delta$ is a clause such that every integer
    term occurring in $\Gamma$ or in $\Delta$ is a
    variable\footnote{Recall that by definition a clause cannot contain arithmetic atoms.}. }
    \end{itemize}
 \end{definition}

The property that in a $\dZ$-clause $\absclause$, every integer term
occurring in $\Gamma$ or in $\Delta$ is a variable is simple to
ensure. If this is not the case, i.e., if $\Gamma, \Delta$ contain an integer
term $t$ that is not a variable, then it suffices to replace every
occurrence of $t$ with a fresh integer variable $u$, and add the
equation $u\iseq t$ to
$\Lambda$. This way
every set of clauses
can be transformed into an equivalent set of $\dZ$-clauses.

The notions of position, replacement, etc.  extend straightforwardly
to sequences of atoms and $\dZ$-clauses, taking them as terms
with $3$ arguments.
The notion of satisfiability is extended to $\dZ$-clauses as follows:

\begin{definition}
  A substitution $\sigma$ is a \emph{solution} of a sequence of
  arithmetic atoms $\Lambda$ in an interpretation $I$ if $\sigma$ maps
  the variables occurring in $\Lambda$ to integers such that $I
  \models \Lambda\sigma$.  A $\dZ$-clause $\absclause$ is \emph{satisfied}
  by an interpretation $I$ if for every solution $\sigma$ of $\Lambda$,
  the clause $(\Gamma \rightarrow \Delta)\sigma$ is satisfied by $I$.
\end{definition}

Note that, although the signature may contain uninterpreted symbols of sort $\dZ$ (e.g. constant
symbols that must be interpreted as integers), it is sufficient to instantiate the integer variables by
integers only.
\begin{definition}
  Given a $\dZ$-clause $C = \absclause$, an \emph{abstraction
    atom} in $C$ is an atom of the form $x\iseq t$ which occurs
  in $\Lambda$. $x \iseq t$
  is \emph{grounding} if $t$ is
  ground. $C$ is \emph{$\dZ$-closed} if all its integer variables
  occur in grounding abstraction atoms and \emph{closed} if it is $\dZ$-closed and every variable
  occurring in $C$ is of sort $\dZ$.
\end{definition}

Intuitively, if $C$ is $\dZ$-closed, this means that $C$ would not
contain any integer variable, had integer terms not been abstracted out.
\articlereport{}{
Abstraction atoms can be viewed  as instantiations, as expressed by
the following proposition:

\begin{proposition}\label{prop_equiv_abstr}
  Given a formula $\phi$ such that $\Var{\phi} = \set{x_1, \ldots,
    x_n}$ and a sequence
    of abstraction atoms $\Lambda = \setof{x_i
    \iseq t_i}{i=1,\ldots, n}$, the sets
  \begin{eqnarray*}
    \exists x_1\cdots \exists x_n.\, (\phi \wedge \Lambda) \textrm{
      and}\\
    \phi \setof{x_i\lfta t_i}{i=1,\ldots, n}
  \end{eqnarray*}
  are equivalent.
\end{proposition}

\begin{proposition}
  Let $C = \absclause$, and assume $\Lambda$ contains an abstraction
  atom $x \iseq s$.  Given $p$, a position of $x$ in $\Gamma
  \rightarrow \Delta$, let
  $C' = C[s]_p$, and let
  $C'' = C\set{x \leftarrow s}$. Then
  $C$, $C'$ and $C''$ are equivalent.
\end{proposition}

\begin{example}
  Let
  \begin{eqnarray*}
    C_1& =& \abscl{x \iseq a, x\iseq b}{}{f(x) \iseq c, g(x) \iseq
      c'},\\
    C_2 &= &\abscl{x\iseq a, x\iseq b}{}{f(x) \iseq c, g(b)\iseq
      c'},\\
    C_3 & =& \abscl{a\iseq a, a\iseq b}{}{f(a) \iseq c, g(a) \iseq c},\\
    C_4 &=& \abscl{a\iseq a, a\iseq b}{}{f(a) \iseq c, g(b)
      \iseq c'}.
  \end{eqnarray*}
  $C_2$ is obtained from $C_1$ by replacing the occurrence of $x$ in
  $g(x)$, i.e. the subterm at position 3.2.1.1 in $C_1$, by constant
  $b$, $C_3$ is obtained by instantiating $C_1$ with the substitution
  $\sigma = \set{x\leftarrow a}$ and $C_4$ is obtained by
  instantiating $C_2$ with $\sigma$. All these clauses are equivalent.
\end{example}
}
We define an operation permitting to add arithmetic atoms to a $\dZ$-clause:

\begin{definition}\label{def_addabs}
  Consider a $\dZ$-clause $C = \absclause$
and a set of arithmetic
  atoms $\Lambda'$. We denote by $\addabs{\Lambda'}{C}$
  the $\dZ$-clause $\abscl{\Lambda', \Lambda}{\Gamma}{\Delta}$.
\end{definition}

\subsubsection*{An inference system for $\dZ$-clauses.}

We denote by $\infsys$ the inference system of \cite{AlthausKW09} on abstracted
clauses, depicted in Figure \ref{fig:zinf}.
Reduction rules are also defined in \cite{AlthausKW09}; the only one that is useful in our context
is the
\emph{tautology deletion} rule also depicted in Figure \ref{fig:zinf}.
\begin{figure}[ht]
  \fbox{
\begin{minipage}{0.9\textwidth}
{\small \begin{description}
\item [Superposition left]:
  $\irule{\abscl{\Lambda_1}{\Gamma_1}{\Delta_1, l\iseq r} ~~~
    \abscl{\Lambda_2}{s[l'] \iseq t, \Gamma_2}{\Delta_2}}
  {(\abscl{\Lambda_1, \Lambda_2}{s[r] \iseq t, \Gamma_1,
      \Gamma_2}{\Delta_1, \Delta_2})\sigma}$\\
  where $\sigma$ is an mgu of $l$ and $l'$, $l\sigma \not\prec
  r\sigma$, $s\sigma\not \prec t\sigma$, $l'$ is not a variable,
  $(l\iseq r)\sigma$ is strictly maximal in $(\Gamma_1 \rightarrow
  \Delta_1, l\iseq r)\sigma$ and $(s[l'] \iseq t)\sigma$ is strictly
  maximal in $(s[l'] \iseq t, \Gamma_2 \rightarrow \Delta_2)\sigma$.
\item [Superposition right]:
  $\irule{\abscl{\Lambda_1}{\Gamma_1}{\Delta_1, l\iseq r} ~~~
    \abscl{\Lambda_2}{\Gamma_2}{\Delta_2, s[l'] \iseq t}}
  {(\abscl{\Lambda_1, \Lambda_2}{\Gamma_1, \Gamma_2}{\Delta_1,
      \Delta_2, s[r] \iseq t})\sigma}$\\
  where $\sigma$ is an mgu of $l$ and $l'$, $l\sigma \not\prec
  r\sigma$, $s\sigma\not \prec t\sigma$, $l'$ is not a variable,
  $(l\iseq r)\sigma$ is strictly maximal in $(\Gamma_1 \rightarrow
  \Delta_1, l\iseq r)\sigma$ and $(s[l'] \iseq t)\sigma$ is strictly
  maximal in $\Gamma_2 \rightarrow \Delta_2, (s[l'] \iseq t)\sigma$.
\item [Equality factoring]: $\irule{\abscl{\Lambda}{\Gamma}{\Delta,
      l\iseq r, l'\iseq r'}} {(\abscl{\Lambda}{\Gamma, r\iseq
      r'}{\Delta, l'\iseq r'})\sigma}$\\
  where $\sigma$ is an mgu of $l$ and $l'$, $l\sigma \not\prec
  r\sigma$, $l'\sigma\not \prec r'\sigma$ and
  $(l\iseq r)\sigma$ is maximal in $(\Gamma_1 \rightarrow
  \Delta_1, l\iseq r, l'\iseq r')\sigma$.
\item [Ordered factoring]: $\irule{\abscl{\Lambda}{\Gamma}{\Delta,
      E_1, E_2}} {(\abscl{\Lambda}{\Gamma}{\Delta, E_1})\sigma}$\\
  where $\sigma$ is an mgu of $E_1$ and $E_2$, and $E_1\sigma$
  is maximal in $(\Gamma\rightarrow \Delta, E_1, E_2)\sigma$.
\item [Equality resolution]: $\irule{\abscl{\Lambda}{\Gamma, s\iseq
      t}{\Delta}}
  {(\abscl{\Lambda}{\Gamma}{\Delta})\sigma}$\\
  where $\sigma$ is an mgu of $s$ and $t$, and $(s\iseq
  t)\sigma$ is maximal in $(\Gamma, s\iseq t \rightarrow
  \Delta)\sigma$.
\item [Constraint refutation]: $\irule{\absclempty{\Lambda_1}\ ~~~
    \cdots ~~~\ \absclempty{\Lambda_n}} {\emptcl}$\\
  where $\absclempty{\Lambda_1}\wedge\cdots \wedge
  \absclempty{\Lambda_n}$ is inconsistent in $\dZ$.
\end{description}
As usual the system is parameterized by an ordering among terms, extended into an ordering on atoms and clauses (see \cite{BG94} for details). The rules are applied modulo the AC properties of the
  sequences
  and the commutativity of $\iseq$.
\begin{description}
\item[Tautology deletion]: $\srule{\absclause}{}$,\\
  if $\Gamma \rightarrow \Delta$ is a tautology, or the existential
  closure of $\Lambda$
  is $\dZ$-unsatisfiable.
\end{description}
}
\end{minipage}
}

\caption{The inference system $\infsys$\label{fig:zinf}}
\end{figure}
We make the additional (and natural) assumption that the ordering is
such that all constants are smaller than all non-flat terms. In order
to obtain a refutational completeness result on this calculus, the
authors of \cite{BachmairGW94,AlthausKW09} impose the condition of
\emph{sufficient completeness} on sets of clauses. Without this
condition, we have the following result, stating a weaker version of
refutational completeness for the calculus.

\begin{theorem}\label{th_sem_complete}
  Let $S$ denote a $\dZ$-unsatisfiable set of $\dZ$-clauses. % Niko add $\dZ$
  Then there exists a
  $\dZ$-unsatisfiable set of clauses $\{ \absclempty{\Lambda_i}\ \mid i \in {\dN} \}$ such
  that for every $i \in {\dN}$, $\absclempty{\Lambda_i}$
  can be deduced from $S$ by applying
  the rules in $\infsys$.
\end{theorem}

\articlereport{}{
  \begin{proof}(Sketch) Let $I$ be an interpretation of the integer
    symbols in $S$, and let $(a_i)_{i \in {\dZ}}$ be a family of
    constant symbols of a new sort $\dZ'$.  We denote by $S'$ the set of
    clauses of the form $C\sigma\normI$ where:
\begin{itemize}
\item{$\myabscl{\Lambda}{C} \in S$.}
\item{$\sigma$ is a substitution mapping every integer variable in $C$
    to a ground integer term $\su^k(0)$ or $-\su^k(0)$ such that $I
    \models \Lambda\sigma$.}
\item{$C\sigma\normI$ is obtained from $C\sigma$ by replacing every ground integer term $t$ by $a_{I(t)}$.}
\end{itemize}
It is clear that $S'$ is a set of clauses ($\dZ$ is replaced by $\dZ'$
in the profile of the function symbols), that contains no integer
term.  Furthermore, $S'$ is unsatisfiable: if $S'$ admits a model $J$,
then a model $K$ of $S$ can be constructed by extending the
interpretation $I$ as follows: for every function $f$ of profile
$\asort_1 \times \ldots \times \asort_n \rightarrow
\asort$ where $\asort \neq \dZ$,
$f^K(d_1,\ldots,d_n) = f^J(d_1',\ldots,d_n')$ where $d_i' = d_i$ if
$\asort_i \not = \dZ$ and $d_i' = J(a_{d_i})$
if $\asort_i = \dZ$. It is
straightforward to check that $K\models S$ if $J \models S'$.

By the refutational completeness of the superposition calculus (on
clauses not containing integer terms), $S'$ admits a refutation. Note
that by definition, no superposition within terms of sort $\dZ'$
 can occur in $S'$.  Thus
by the usual lifting argument, this derivation can be transformed into
a derivation from $S$: it suffices to replace in the clauses $C\sigma$
each constant $a_{i}$ by the corresponding integer variable in $C$ and
to attach the original arithmetic constraint $\Lambda$ to the clause.
This derivation yields a clause of the form $\absclempty{\Lambda}{}$
where $I \models \exists \vec{x}.\Lambda$ (if $\vec{x}$ denotes the
vector of variables in $\Lambda$).  By repeating this for every
possible interpretation, we obtain a set of clauses satisfying the
desired property.  The conjunction of these clauses is unsatisfiable
since every arithmetic interpretation falsifies at least one clause.
\end{proof}
}

Note that this does \emph{not}
imply refutational completeness, since the set
$\{ \absclempty{\Lambda_i} \mid i \in {\dN} \}$ may be infinite (if
this set is finite then the \textbf{Constraint refutation} rule
applies and generates $\Box$). For instance, the set of $\dZ$-clauses
$S = \{ \abscl{x \iseq a}{p(x)}{},\ \abscl{x \iseq s(y)}{p(x)}{p(y)},\
p(0),\ \absclempty{a < 0} \}$ is clearly unsatisfiable, and the
calculus generates an infinite number of clauses of the form
$\absclempty{\su^k(0) \iseq a}$, for $k \in {\dN}$.  It is actually
simple to see that there is no refutationally complete calculus for
sets of $\dZ$-clauses, since we explicitly assume that $\dZ$ is
interpreted as the set of integers. In our case however there are
additional conditions on the arithmetic constraints that ensure that
only a finite set of $\dZ$-clauses of the form $\absclempty{\Lambda}$ will be generated. Thus, for the
$\dZ$-clauses we consider, refutational completeness of the calculus
will hold, and it will always generate the empty clause
starting from an unsatisfiable set of $\dZ$-clauses.
However,
we do not intend to use this result
 to test the satisfiability
of the formulas. The reason is that -- as explained in the
introduction -- the superposition calculus is not well adapted to
handle efficiently very large propositional formulas. In this paper,
we use the inference system $\infsys$
 only as a theoretical tool to show
the \emph{existence} of an instantiation scheme. To this aim we need the
following property (see \cite{BachmairGW94}, Lemma 8 for details):

\begin{proposition}\label{prop_intvar}
  If $\sigma$ is an mgu occurring in a $\infsys$-inference, then $\sigma$ maps
  integer variables to integer variables.
\end{proposition}

\section{Instantiation of inequality formulas}

Given an SMT problem over a combination of a given theory with the
theory of linear arithmetic, the inference system of
\cite{AlthausKW09} permits to separate the reasoning on the theory
itself from the reasoning on the arithmetic part of the formula. If
the input set of clauses is unsatisfiable, then the inference system
will generate a set of clauses
of the form $\{ \abscl{\Lambda_1}{}{},
\ldots, \abscl{\Lambda_n}{}{},\ldots \}$,
which is inconsistent in
$\dZ$. In
this section, we investigate how to safely instantiate the
$\Lambda_i$'s, under some condition on the atoms they contain. We
shall impose that each $\Lambda_i$ be equivalent to a formula of the
following form:

\begin{definition}
An \emph{inequality formula} is of the form
$\phi:\ \bigwedge_{i=1}^m s_i \isless t_i$,
where for all $i = 1,\ldots,m$, $s_i$  and $t_i$ are ground terms or variables.
\end{definition}

If $A$ is a set of terms, we use the notation $A \isless x$ (resp. $x\isless
A$) as a shorthand for $\bigwedge_{s\in A}s \isless x$
(resp. $\bigwedge_{s\in A} x \isless s$). We
denote by $U^\phi_x$ the set
$U^\phi_x\ =\ \setof{y \in \Var{\phi}}{x \isless y \textrm{
    occurs in } \phi}$.
We may thus rewrite the formula $\phi$ as
\articlereport{$\bigwedge_{x \in \Var{\phi}}(A_{x}^{\phi} \isless x \wedge x
\isless \upbnd{x}{\phi} \wedge \bigwedge_{y \in U^\phi_{x}}x \isless y)
\wedge \psi$,}{
\[\phi: \bigwedge_{x \in \Var{\phi}}(A_{x}^{\phi} \isless x \wedge x
\isless \upbnd{x}{\phi} \wedge \bigwedge_{y \in U^\phi_{x}}x \isless y)
\wedge \psi,\]} where the sets $A_{x}^{\phi}$ and $\upbnd{x}{\phi}$ are
ground for all $x$, and $\psi$ only contains inequalities between
ground terms.

\begin{definition}
  For all $x \in \Var{\phi}$, we consider the sets
  $\mybsup{B}{x}{\phi}$, defined as the smallest sets satisfying \articlereport{$\mybsup{B}{x}{\phi}\ \supseteq\ \upbnd{x}{\phi}\
    \cup\ \bigcup_{y \in U^\phi_x}\mybsup{B}{y}{\phi} \cup
  \set{\speconst}$,}{the following property:
  \[\mybsup{B}{x}{\phi}\ \supseteq\ \upbnd{x}{\phi}\ \cup\ \bigcup_{y
    \in U^\phi_x}\mybsup{B}{y}{\phi} \cup
  \set{\speconst},\]}
  where $\speconst$ is a special constant that does not occur in
  $\phi$.
\end{definition}

\articlereport{}{
Note that the $\mybsup{B}{x}{\phi}$'s are never empty.
}
\articlereport{}{
\begin{example}
  Consider the formula $\phi$:
  \[\begin{array}{ccccccc}
    a_1\isless x &\wedge& x\isless b_1& \wedge& x\isless y &\wedge &
    x\isless z\\
    a_2 \isless y& \wedge& y \isless b_2 &\wedge &y\isless b_3&\wedge
    & a_3\isless z.
  \end{array}\]
  Then $\mybsup{B}{x}{\phi} = \set{b_1, b_2, b_3,\speconst}$,
  $\mybsup{B}{y}{\phi} = \set{b_2, b_3, \speconst}$, and
  $\mybsup{B}{z}{\phi} = \set{\speconst}$.
\end{example}
}

\articlereport{}{
\begin{proposition}
\label{prop_for_th14}
For every inequality formula $\phi$,  every variable $x$ occurring
in $\phi$ and for every term $t \in \mybsup{B}{x}{\phi}$, either $t =
\chi$ or $\phi \models_{\dZ} x \isless t$.
\end{proposition}

\begin{proof}
  Let $C_x$ be the set of terms $t$ such that either $t = \chi$ or
  $\phi \models_{\dZ} x \isless t$.  By definition, $C_x$ contains
  $B_x^\phi$ and $\chi$.  Furthermore, if $y \in U_x^\phi$ then $x
  \isless y$ occurs in $\phi$ thus $\phi \models x \isless y$ hence
  $\phi \models x \isless t$, for every $t \in C_y$ distinct from
  $\chi$ (by transitivity of $\leq$ in $\dZ$).  Thus $C_x \supseteq
  \bigcup_{y \in U_x^\phi} C_y$ and finally $C_x \supseteq B_x^\phi
  \cup \bigcup_{y \in U_x^\phi} C_x \cup \{ \chi \}$.  By definition
  the $\mybsup{B}{x}{\phi}$'s are the smallest sets satisfying the
  previous property, thus $\forall x \in \Var{\phi}, C_x^\phi
  \supseteq \mybsup{B}{x}{\phi}$ and the proof is completed.
\end{proof}
}

\begin{theorem}\label{th_instz}
  Given an inequality formula $\phi$ such that $\Var{\phi} = \set{x_1,
    \ldots, x_n}$
  consider the two following formulas:
  \[
  \begin{array}{cr}
    [\exists x_1 \cdots \exists x_n.\phi] & (\alpha)\\
    \left(\bigvee_{s_1 \in \mybsup{B}{x_1}{\phi}}\cdots \bigvee_{s_n \in \mybsup{B}{x_n}{\phi}}
    \phi \setof{x_i\leftarrow s_i}{i=1,\ldots, n} \right)  & (\beta)
  \end{array}
  \]
  %where   constant $\speconst$ does not
  %appear in $(\alpha)$.
  Let $I$ denote a $\dZ$-interpretation of $(\alpha)$
 and $G$ denote a ground set containing all ground terms occurring in $\phi$. Then $I \modelz (\alpha)$ if and only if, for any
  extension $J$ of $I$ to the constant $\speconst$, $J \modelz
  \left(\bigwedge_{t\in G}\neg(\speconst \isless t)\right )
  \Rightarrow (\beta)$.
%$J \modelz
%  \left(\bigwedge_{t\in B_{x_i}^{\phi}, i \in [1..n]}\neg(\speconst \isless t)\right )
%  \Rightarrow (\beta)$.
\end{theorem}

\articlereport{}{
\begin{proof}
 First suppose all extensions of $I$ to $\speconst$ satisfy
 $(\beta)$, and let $J$ denote an extension of $I$ such that for all
 $t \in G$, $J(\speconst) > J(t)$. Then by construction of $(\beta)$,
 there exists a substitution $\sigma = \setof{x_i \lfta
   s_i}{i=1\ldots, n}$, where for all $i = 1, \ldots, n$, $s_i \in
 \mybsup{B}{x_i}{\phi}$, such that $J \modelz \phi\sigma$. It is clear
 that $I \modelz \phi\sigma$ since $\speconst$ does not occur in
 $\phi$.

 Conversely, assume that $I \models_{\dZ} (\alpha)$.  Let $J$ be an
 extension of $I$ such that $\forall t \in G, J(\chi) > J(t)$.  Let
 $\sigma$ be the substitution of domain $\{ x_1,\ldots,x_n \}$ such
 that for all $i \in [1..n]$, $\sigma(x_i) = \min_{t \in
   \mybsup{B}{x_i}{\phi}}(I(t))$. We prove that $J \models
 \phi\sigma$.

$I \models \exists x_1\ldots\exists x_n.\phi$, thus there exists a
substitution $\theta$ mapping each variable $x_i$ $(1 \leq i \leq n$)
to an integer such that $I \models \phi\theta$.  For every atom $t
\isless s$ occurring in $\phi$, we have $I \models (t \isless
s)\theta$. We prove that $J \models (t \isless s)\sigma$ by
investigating the different cases:

\begin{itemize}
\item{If $t,s$ are two ground terms in $\phi$, then since $\chi$ does
    not occur in $\phi$, we have $I(t) = J(t)$ and $I(s) = J(s)$ thus
    $I \models (t \isless s) \Rightarrow J \models (t \isless s) = (t
    \isless s)\sigma$.}
\item{
If $t$ is a variable $x_i$ (for some $i \in [1..n]$) and $s$ is a ground term,
then by definition $s \in B_{x_i}^\phi \subseteq \mybsup{B}{x_i}{\phi}$, thus by definition of $\sigma$, we have
$x_i\sigma \leq I(s) = J(s)$. Hence $J \models (t \isless s)\sigma$.}

\item{ If $s$ is a variable $x_i$ (for some $i \in [1..n]$) and $t$ is
    a ground term, then by Proposition \ref{prop_for_th14}, since
    $x_i\sigma \in I(\mybsup{B}{x_i}{\phi})$ we must have either $\phi
    \models x_i \isless x_i\sigma$ or $x_i\sigma = I(\chi)$.  If
    $x_i\sigma = I(\chi)$ then since $t$ is in $G$, we have
%     $t$ occurs in $\bigcup_{i \in [1..n]} B_{x_i}^{\phi}$ and we have
      $J \modelz (t \isless x_i\sigma)$ by
     definition of $J$.
Otherwise, by definition $\phi \modelz t
    \isless x_i$, thus $\phi \modelz t \isless x_i\sigma$, by
    transitivity of $\leq$ in $\dZ$. Hence $I \modelz (t \isless
    x_i)\sigma$, i.e. $J \modelz (t \isless x_i)\sigma$.}
\item{
Finally, assume that both $t$ and $s$ are variables $x_i,x_j$ respectively, where $i,j \in [1..n]$.
Since $x_i \isless x_j$ occurs in $\phi$ we have $\mybsup{B}{x_i}{\phi} \supseteq \mybsup{B}{x_j}{\phi}$.
Thus $\min I(\mybsup{B}{x_i}{\phi}) \leq \min I(\mybsup{B}{x_j}{\phi})$, hence $x_i\sigma \leq x_j\sigma$ i.e. $J \models (t \isless s)\sigma$.
}
\end{itemize}
\end{proof}
}

In our case, the sets $B_{x_i}^\phi$ will not be known,
since the clauses in which $\phi$ occurs will not be generated explicitly (see Section \ref{sec_instz} for details). Thus we need to
use an \emph{over-approximation} of these sets:

\begin{definition}\label{def_ub}
  A set of ground terms $B$ is an \emph{upper bound} of an
  inequality formula $\phi$ if for all atoms $x\isless t$ occurring in
  $\phi$, $t$ is an element of $B$. The set $B$ is an \emph{upper
    bound} of a set of inequality formulas if it is an upper bound of
  each formula.
\end{definition}

\articlereport{}{
\begin{proposition}
Let $\phi$ denote an inequality formula.
If $B$ is an upper bound of $\phi$ then for every variable $x$ in $\phi$,
$B \cup \{ \chi \} \supseteq \mybsup{B}{x}{\phi}$.
\end{proposition}
}

It is clear that if $B$ is an upper bound of an inequality formula,
then Theorem \ref{th_instz} still holds when the variables in $\phi$
are instantiated by all the elements in $B \uplus \set{\speconst}$
instead of just those in the $\mybsup{B}{x}{\phi}$'s.

\begin{definition}
  Given an inequality formula $\phi$ such that $\Var{\phi} = \set{x_1,
    \ldots, x_m}$ and a set of ground terms $B$, a \emph{{\bdef} of
    $\phi$} is a set (i.e. a conjunction) of grounding abstraction
  atoms $\setof{x_i \iseq s_i}{i=1, \ldots, m}$, such that every
  $s_i$ is in $B$.  We denote by $\setbdefs{B}{\phi}$ the set of all
  {\bdef}s of $\phi$.
\end{definition}

Intuitively, a {\bdef} of a formula represents a grounding
instantiation of this formula.
\articlereport{}{
\begin{example}
  Let $\phi = x\isless f(a) \wedge b \isless x \wedge y\isless a$, and assume
  $B = \set{a,c}$. Then $\setbdefs{B}{\phi}$ contains four sets:
  \begin{itemize}
  \item $\set{x\iseq a, y\iseq a}$, which corresponds to substitution
    $\set{x\lfta a, y\lfta a}$,
  \item $\set{x\iseq a, y\iseq c}$, which corresponds to substitution
    $\set{x\lfta a, y\lfta c}$,
  \item $\set{x\iseq c, y\iseq a}$, which corresponds to substitution
    $\set{x\lfta c, y\lfta a}$,
  \item $\set{x\iseq c, y\iseq c}$, which corresponds to substitution
    $\set{x\lfta c, y\lfta c}$.
  \end{itemize}
\end{example}
}
We rephrase a direct
consequence of Theorem \ref{th_instz} using {\bdef}s:

\begin{corollary}\label{cor_instz}
  Let $\set{\phi_1, \ldots, \phi_n}$ denote a set of inequality
  formulas over the disjoint sets of variables $\setof{x_{i,1},
    \ldots, x_{i,m_i}}{i=1\ldots, n}$, let $B$ denote an upper bound
  of this set, and assume that $\bigvee_{i=1}^n \exists x_{i,1} \cdots
  \exists x_{i, m_i}.\,\phi_i$ is valid in $\dZ$. If $G$contains all ground terms occurring in the inequality formulas and $B' = B \uplus
  \set{\speconst}$, then
  \[\bigwedge_{t\in G} \neg(\speconst \isless t) \Rightarrow
  \bigvee_{i=1}^n \left(\bigvee_{\Lambda_i' \in
      \setbdefs{B'}{\phi_i}}\exists x_{i,1}\cdots \exists x_{i, m_i}.\, \phi_i\wedge \Lambda_i'\right)\ \ \ \ \ \ (\gamma),\]
  is also valid in $\dZ$.
\end{corollary}

\articlereport{}{
\begin{proof}
  Let $I$ denote a $\dZ$-interpretation that interprets all the
  symbols in $\Sigma \uplus \set{\speconst}$; by hypothesis, $I$ must
  satisfy the formula $\bigvee_{i= 1}^n \exists x_{i,1} \cdots \exists
  x_{i, m_i}\phi_i$ in $\dZ$. Since for all $i = 1, \ldots, n$ and for
  all terms $s_{i,1}, \ldots, s_{i,m}$, the formulas
  \[\begin{array}{c}
  \phi_i\setof{x_{i,j} \lfta s_{i,j}}{j = 1, \ldots, m_i} \textrm{
    and}\\
  \exists x_{i,1}\cdots \exists x_{i, m_i}.\, \phi_i \wedge \bigwedge_{j = 1}^{m_i} x_{i,j} \iseq s_{i,j}
\end{array}\]
are equivalent by Proposition \ref{prop_equiv_abstr}, we deduce by
Theorem \ref{th_instz} that
$I$ is also a
model of $(\gamma)$, hence the result.
\end{proof}
}

It is important to note that results similar to those of this section
could have been proved by considering the terms occurring in atoms of
the form $t\isless x$, instead of those of the form $x\isless t$, and
considering lower bounds instead of upper bounds. This should allow to
choose, depending on the problem and which sets are smaller, whether
to instantiate variables using lower bounds or upper bounds.

\section{Properties of inferences on $\dZ$-clauses}\label{sec_instz}

Corollary \ref{cor_instz} shows how to safely get rid of integer
variables in a set of inequality formulas, provided an upper bound of
this set is known. The goal of this section is first to show that given an
initial set of $\dZ$-clauses $S$, such an upper bound can be
determined, regardless of the inequality formulas that can be
generated from $S$. Then we show that by instantiating the integer
variables in $S$, it is still possible to generate all necessary
instances of the inequality formulas. Thus, $S$ and the corresponding
instantiated set will be equisatisfiable.

\articlereport{}{We shall use Proposition \ref{prop_intvar} to describe
several properties on $\dZ$-clauses that are preserved by
inferences.}
We first define a generalization of the notion of an upper bound of an
inequality formula (see Definition \ref{def_ub}), to the case of
$\dZ$-clauses.

\begin{definition}
  Given a $\dZ$-clause $C = \absclause$ and a set of ground terms $B$, we
  write $C \spsub B$ if for all atoms $x \isless t \in \Lambda$,
  \begin{itemize}
  \item $\Lambda$ contains (not necessarily distinct) grounding
    abstraction atoms of the form $x_i \iseq s_i$, $i=1, \ldots,
    n$;
  \item there exist variable positions
    $\set{p_1, \ldots, p_n}$ such that variable $x_i$ occurs at
    position $p_i$, and $t[s_1]_{p_1}\ldots[s_n]_{p_n} \in B$.
\end{itemize}
\end{definition}

\begin{example}
  Let $C = \abscl{x\iseq a, y\iseq b, y\iseq c, z \isless
    f(g(x,y),y)}{}{h(x,y,z) \iseq d}$ and $B = \set{f(g(a,c), b)}$. Then $C \spsub B$.
\end{example}

Intuitively, for a $\dZ$-clause $C = \absclause$, the set $B$ is an upper
bound of the inequality atoms in $\Lambda$ provided for all
atoms $x\isless t$, the variables in $t$ are replaced by the correct
terms.
\articlereport{}{
This property is preserved by inferences:

\begin{lemma}\label{lm_spsub}
  Let $D, D'$ denote (not necessarily distinct) $\dZ$-clauses, that generate
  a $\dZ$-clause $C$, and let $B$ denote a set of ground terms. If $D \spsub
  B$ and $D' \spsub B$, then $C\spsub B$.
\end{lemma}
\begin{proof}
  Let $D = \abscl{\Lambda_1}{\Gamma_1}{\Delta_1}$ and $D' =
  \abscl{\Lambda_2}{\Gamma_2}{\Delta_2}$. Then $C$ is of the form
  $(\abscl{\Lambda_1, \Lambda_2}{\Gamma}{\Delta})\sigma$, where
  $\sigma$ maps integer variables to integer variables by Proposition
  \ref{prop_intvar}. Let $x \isless t$ denote an atom in $(\Lambda_1,
  \Lambda_2)\sigma$, then one of $\Lambda_1, \Lambda_2$, say
  $\Lambda_1$, contains an atom $y\isless t'$, such that $(y\isless
  t')\sigma = x \isless t$. Since $D\spsub B$, by hypothesis, $\Lambda_1$
  contains grounding abstraction atoms of the form $x_i\iseq s_i$
  and there exists variable positions $\set{p_1, \ldots, p_n}$ such
  that $t'[s_1]_{p_1}\ldots[s_n]_{p_n} \in
  B$. It is clear that for all $i$, $(x_i \iseq s_i)\sigma$ is a
  grounding abstraction atom in $\Lambda_1\sigma$, thus
  $t[s_1]_{p_1}\ldots[s_n]_{p_n} =
  t'[s_1]_{p_1}\ldots[s_n]_{p_n} \in B$.
\end{proof}
}
In order not to unnecessarily instantiate some of the integer
variables in a $\dZ$-clause, we distinguish those that appear in abstraction
atoms from those that appear in inequality atoms. It will only
be necessary to instantiate the latter variables.

\begin{definition}
  Let $C=\abscl{\Lambda}{\Gamma}{\Delta}$; the
 set of \emph{abstraction variables in $C$} $\absvar{C}$ and the set of
 \emph{inequality variables in $C$} $\ineqvar{C}$ are defined as follows:
  \articlereport{$\absvar{C}\ =\ \setof{x \in \Var{C}}{\Lambda \textrm{ contains an
      abstraction atom } x\iseq t}$}{
  \[\absvar{C}\ =\ \setof{x \in \Var{C}}{\Lambda \textrm{ contains an
      abstraction atom } x\iseq t}\]} and
  \articlereport{$\ineqvar{C}\ =\ \setof{x \in \Var{C}}{\Lambda \textrm{ contains an
      atom of the form } x\isless t \textrm{ or } t \isless x}$}{\[\ineqvar{C}\ =\ \setof{x \in \Var{C}}{\Lambda \textrm{ contains a
      atom of the form } x\isless t \textrm{ or } t \isless x}.\]}
\end{definition}
We may assume without loss of generality that all integer variables in
a $\dZ$-clause $C$ are  in $\absvar{C} \cup \ineqvar{C}$. If this
is not the case, it suffices to add to the arithmetic part of $C$ the
atom $x\isless x$.

We define the notion of a \emph{{\preconstrained} $\dZ$-clause}. If a
{\preconstrained} $\dZ$-clause is of the form $\abscl{\Lambda}{}{}$, then
$\Lambda$ will be equivalent to an inequality formula, and this
property is preserved by inferences.

\begin{definition}
  A $\dZ$-clause $C = \absclause$ is \emph{\preconstrained} if every atom
  in $\Lambda$ that is not a grounding abstraction atom either has
  all its variables in $\absvar{C}$, or is of the form $x\isless t$ or
  $t\isless x$, where $t$ is either a variable itself, or has all its
  variables  in $\absvar{C}$.
\end{definition}

\begin{example}

 $\abscl{x\iseq a, y \iseq b, f(x,y) \iseq g(y), z \isless
      g(x)}{}{h(x,y,z) \iseq e}$ is {\preconstrained} but $\abscl{x\iseq a,  y \isless g(y)}{}{h(x,y,z) \iseq e}$ is not because $y$ does not
  occur in a grounding abstraction atom.
\end{example}

\articlereport{}{
\begin{lemma}\label{lm_pres_prec}
  Let $D, D'$ denote (not necessarily distinct) $\dZ$-clauses that generate
  a $\dZ$-clause $C$. If $D$ and $D'$ are {\preconstrained}, then so is $C$.
\end{lemma}

\begin{proof}
  This is a direct consequence of Proposition \ref{prop_intvar}, since
  integer variables are mapped to integer variables by mgu $\sigma$,
  and it is simple to verify that $\absvar{C} = (\absvar{D}
  \cup \absvar{D'})\sigma$.
\end{proof}
}

\articlereport{}{
\begin{definition}
  Given a $\dZ$-clause $C = \absclause$, we denote by $\absterm{C}$ the set
  \begin{eqnarray*}
    \absclause & = & \setof{t}{x\iseq t \textrm{ is a grounding
        abstraction atom in }\Lambda}.
  \end{eqnarray*}
  Given a set of $\dZ$-clauses $S$, we denote by $\absterm{S}$
  the set $\bigcup_{C \in S}\absterm{C}$.
\end{definition}
}
\articlereport{}{
\begin{lemma}\label{lm_absterm_inc}
  If $C$ is generated from a set of clauses $S$, then $\absterm{C}
  \subseteq \absterm{S}$.
\end{lemma}

\begin{proof}
  This is a direct consequence of Proposition \ref{prop_intvar}.
\end{proof}
}

We extend the notion of a {\bdef} to
$\dZ$-clauses. Intuitively, a {\bdef} of such a $\dZ$-clause
represents a ground instantiation of the inequality variables it contains.

\begin{definition}
  Given a $\dZ$-clause $C$ such that $\ineqvar{C} = \set{x_1, \ldots, x_m}$
  and a set of ground terms $B$, a \emph{{\bdef} of $C$} is a set of
  grounding abstraction atoms $\setof{x_i \iseq s_i}{i=1, \ldots,
    m}$, such that every $s_i$ is in $B$.  We
  denote by $\setbdefs{B}{C}$ the set of all {\bdef}s of $C$.
    Given a set of $\dZ$-clauses $S$, we denote by $\addsetabs{S}{B}$ the set
  $\addsetabs{S}{B}\ =\ \setof{\addabs{\Lambda'}{C}}{C \in S \wedge
    \Lambda' \in \setbdefs{B}{C}}$.
\end{definition}

\articlereport{}{
If a $\dZ$-clause $C$ is generated from a set of $\dZ$-clauses $S$, the following
lemma shows that by carefully instantiating the inequality variables
occurring in $S$, we obtain a set of $\dZ$-clauses that generates the
required instances of $C$.

\begin{lemma}
  Consider a set of $\dZ$-clauses $S$ and  a set of terms $B$ such that
  $\absterm{S} \subseteq B$. If $C = \absclause$ is generated from $S$ and
  $\Lambda' \in \setbdefs{B}{C}$, then $C' = \addabs{\Lambda'}{C}$ can be
  generated from $\addsetabs{S}{B}$.
\end{lemma}

\articlereport{}{
\begin{proof}
  By induction on the length of the derivation. Assume $C$ is
  generated by a superposition from $D_1 =
  \abscl{\Lambda_1}{\Gamma_1}{\Delta_1}$ into $D_2 =
  \abscl{\Lambda_2}{\Gamma_2}{\Delta_2}$; the other cases are
  identical. Then $C$ is of the form $(\abscl{\Lambda_1,
    \Lambda_2}{\Gamma}{\Delta})\sigma$, and $\sigma$ maps the
  variables in $\Lambda = \Lambda_1\cup \Lambda_2$ to variables. For
  $i = 1, 2$, let
  \[\Lambda_i'\ =\ \setof{x \iseq t}{x \in \ineqvar{D_i} \textrm{ and
    } x\sigma \iseq t \textrm{ is an abstraction atom in } (\Lambda \cup
    \Lambda')\sigma},\]
  Then $C'$ is generated by $D_1' = \addabs{\Lambda_1'}{D_1}$ and
  $D_2' = \addabs{\Lambda_2'}{D_2}$.

  By Lemma \ref{lm_absterm_inc}, $\absterm{C} \subseteq \absterm{S}$,
  hence $\absterm{C'} \subseteq B$. Thus, for $i =1,2$, $\Lambda_i'
  \in \setbdefs{B}{D_i}$; by the induction hypothesis, $D_1'$ and
  $D_2'$ are generated by $\addsetabs{S}{B}$, therefore, so is $C'$.
\end{proof}
}

\begin{example}
  Consider the $\dZ$-clauses $D_1 = \abscl{x\iseq i}{}{\select(a,x) \iseq
    e}$ and $D_2 =
  \abscl{y\isless b}{}{\select(a,y) \iseq e'}$, and let $B =
  \set{i}$. These $\dZ$-clauses generate $C = \abscl{x\iseq i, x \isless
    b}{}{e \iseq e'}$, which is also generated by $D_1$ and $\addabs{y
    \iseq i}{D_2}$.
\end{example}

The following relation permits to keep track of the ground integer
terms that may occur in a derivation:

\begin{definition}
   Given a $\dZ$-clause $C = \absclause$ and a set of ground terms $G$, we
  write $C \spinc B$ if for all nonvariable terms $t$ in $\Lambda$,
  \begin{itemize}
  \item $\Lambda$ contains (not necessarily distinct) grounding
    abstraction atoms of the form $x_i \iseq s_i$, $i=1, \ldots,
    n$;
  \item there exist variable positions
    $\set{p_1, \ldots, p_n}$ such that variable $x_i$ occurs at
    position $p_i$, and $t[s_1]_{p_1}\ldots[s_n]_{p_n} \in G$.
\end{itemize}
\end{definition}

We prove a stability result on the set of ground integer terms that
may occur in a derivation:

\begin{proposition}
Let $D, D'$ denote (not necessarily distinct) $\dZ$-clauses, that generate
  a $\dZ$-clause $C$, and let $G$ denote a set of ground terms. If $D \spinc
  G$ and $D' \spinc G$, then $C\spinc G$.
\end{proposition}

We may now state a result which links the
\textbf{constraint refutation} rule of the inference system with
Corollary \ref{cor_instz}, and suggests a way of safely instantiating
inequality variables in a set of $\dZ$-clauses.

\begin{lemma}\label{lm_const_ref}
  Let $B$ denote a set of ground terms, and let $S= \set{C_1, \ldots,
    C_n}$ denote a set of $\dZ$-clauses such that for all $i = 1, \ldots, n$,
  $C_i = \absclempty{\Lambda_i}$ is a {\preconstrained} $\dZ$-clause
  such that $C_i \spsub B$. Given a constant symbol $\speconst$ that
  does not occur in $S$, let $B' = B \cup \set{\speconst}$. If $G
  \supseteq B$ is
  a ground set such that for all $î$, $C_i \spinc G$, then $S$
  is satisfiable in $\dZ$ if and only if
  \[\bigcup_{i=1}^n \setof{\addabs{\Lambda_i'}{C_i}}{\Lambda_i' \in
    \setbdefs{B'}{C_i}} \cup \bigcup_{t\in G} \set{\abscl{\speconst \isless
      t}{}{}}\] is satisfiable in $\dZ$.
\end{lemma}

\articlereport{}{
\begin{proof}
  For $i = 1, \ldots, n$, let $\set{x_{i,1}, \ldots, x_{i, m_i}}$
  denote the set of variables occurring in $\Lambda_i$.  Since
  $\set{C_1, \ldots, C_n}$ is unsatisfiable in $\dZ$, the formula
  $\bigvee_{i=1}^n \exists x_{i,1}\cdots \exists x_{i, m_i}\Lambda_i$
  must be valid in $\dZ$.  Since every $C_i$ is {\preconstrained} and
  such that $C_i \spsub B$, every
  $\Lambda_i$ is equivalent to an inequality formula of the form
  \[\phi_i \equiv \bigwedge x_j \isless s_j' \wedge \bigwedge s_k' \isless x_k \wedge
  \psi',\] over the set of variables $\set{x_{i,1}, \ldots, x_{i,
      m_i}}$, and $\phi_i$ is upper bounded by $B$.

  By Corollary \ref{cor_instz} the formula
  \[\bigwedge_{t\in G} \neg(\speconst \isless t) \Rightarrow
  \bigvee_{i=1}^n \left(\bigvee_{\Lambda_i' \in
      \setbdefs{B'}{\phi_i}}\exists x_{i,1}\cdots \exists x_{i, m_i}.\, \phi_i\wedge \Lambda_i'\right)\]
  is valid in
  $\dZ$ if and only if the formula
  \[\bigwedge_{t\in G}\neg (\speconst \isless t)\
  \Rightarrow\ \bigvee_{i=1}^n \left(\bigvee_{\Lambda_i' \in
      \setbdefs{B'}{C_i}} (\exists x_{i,1}\cdots \exists x_{i, m_i}.\,
    \Lambda_i \wedge \Lambda_i')\right )\]
  is valid in $\dZ$, hence the result.
\end{proof}
}
}
\articlereport{We obtain the main result of this section:}{
We therefore obtain the main result of this section:
}

\articlereport{
\begin{theorem}\label{th_inst}
   Suppose $S$ is a set of $\dZ$-clauses and $B$ is a set of ground terms
  such that for every $\dZ$-clause of the form $C =
  \absclempty{\Lambda}$ generated from $S$, $C$ is {\preconstrained} and $C \spsub B$.
  Let $B' = B \cup \set{\speconst}$, where $\speconst$ does not occur
  in $S$. Then there exists a set of ground terms $G$ containing $B$ and all ground terms of sort $\dZ$ in $S$
  such that $S$ is $\dZ$-satisfiable if and only if $\addsetabs{S}{B'} \cup
  \bigcup_{t \in G}\set{\abscl{\speconst\isless t}{}{}}$ is
  $\dZ$-satisfiable.
\end{theorem}
}{
\begin{theorem}\label{th_inst}
  Suppose $S$ is a set of $\dZ$-clauses and $B$ is a set of ground terms
  such that for every $\dZ$-clause of the form $C =
  \absclempty{\Lambda}$ generated from $S$:
  \begin{itemize}
  \item $C$ is {\preconstrained},
  \item $C \spsub B$.
  \end{itemize}
  Let $B' = B \cup \set{\speconst}$, where $\speconst$ does not occur
  in $S$, and let $G \supseteq B$ denote a set of ground terms such that for all
  $C \in S$, $C \spinc G$. Then $S$ is $\dZ$-satisfiable if and only if $\addsetabs{S}{B'} \cup
  \bigcup_{t \in G}\set{\abscl{\speconst\isless t}{}{}}$ is
  $\dZ$-satisfiable.
\end{theorem}
}
\articlereport{}{
\begin{proof}
  If $S$ is unsatisfiable, then by Theorem
  \ref{th_sem_complete}, it generates an unsatisfiable set of $\dZ$-clauses
  $\setof{C_i}{i \in \dN}$, and each $C_i$ is of the form
  $\absclempty{\Lambda_i}$. The number of such $\dZ$-clauses $S$ can
  generate is finite, since only a finite number of terms can appear
  in the arithmetic part of each $\dZ$-clause, hence
  the \textbf{constraint refutation} rule can be applied to generate
  the empty $\dZ$-clause. By Lemma \ref{lm_const_ref}, $\addsetabs{S}{B'}
  \cup \bigcup_{t \in G}\set{\abscl{\speconst\isless t}{}{}}$ is
  unsatisfiable. Now assume $S$ is satisfiable, then since $\speconst$
  does not occur in $S$, it is clear that $S \cup \bigcup_{t \in
    G}\set{\abscl{\speconst\isless t}{}{}}$ is also satisfiable,
  hence, so is the instantiated set $\addsetabs{S}{B'} \cup \bigcup_{t
    \in G}\set{\abscl{\speconst\isless t}{}{}}$.
\end{proof}
}
In particular, since we may assume all the integer variables occurring
in $S$ are in a $\absvar{C} \cup \ineqvar{C}$ for some $C \in S$,
every $\dZ$-clause occurring in $\addsetabs{S}{B'}$ can be
reduced to a  $\dZ$-clause that is $\dZ$-closed, and
$\addsetabs{S}{B'} \cup \bigcup_{t \in G}\set{\abscl{\speconst\isless
    t}{}{}}$ can be reduced to a set of clauses containing no integer
variable. Hence, Theorem \ref{th_inst} provides a way of getting rid
of all integer variables in a formula.

The instantiated set $\addsetabs{S}{B'} \cup \bigcup_{t \in
  G}\set{\abscl{\speconst\isless t}{}{}}$ can further be reduced:
since  $\chi$ is strictly greater than any ground term $t$
occurring in $S$ or in $B$, % Niko 3
every atom of the form $\chi \isless t$ or $t \isless \speconst$ can be
replaced by false and true respectively.
Furthermore, by construction
$\chi$ only appears at the root level in the arithmetic terms. Thus we
can safely assume that $\chi$ does not occur in the arithmetic part of the
$\dZ$-clause in $S_{B'}$.
This implies that the
inequations $\absclempty{\chi \isless t}$ for $t \in G$ are useless and can be removed. % Niko 3
Note that the resulting set does not depend on $G$.

\section{Completeness of the combined instantiation schemes}\label{sec_comp}

The aim of this section is to determine sufficient conditions
guaranteeing that once the integer variables have been instantiated,
another instantiation scheme can be applied to get rid of the
remaining variables in the set of clauses under consideration.

Let $\aclass$ denote a class of clause sets admitting an instantiation
scheme, i.e., a function $\instgamma$ that maps every clause set $S
\in \aclass$ to a finite set of ground instances of clauses in $S$,
such that $S$ is satisfiable if and only if $\instgamma(S)$ is
satisfiable. If $\instgamma(S)$ is finite, this implies that the
satisfiability problem is decidable for $\aclass$.  For every clause
$C$ in a set $S \in \aclass$, we denote by $\instcl{S}{C}$ the set of
ground substitutions $\sigma$ such that $C\sigma \in \instgamma(S)$.
Thus, by definition, $\instgamma(S) = \{ C\instcl{S}{C} \mid C \in S
\}$.  Since $\instgamma$ is generic, we do not assume that it
preserves $\dZ$-satisfiability. In order to apply it in our
setting, we need to make additional assumptions on the instantiation
scheme under consideration.

\begin{definition}\label{def_instadmin}
  A term $t$ is \emph{\disc}\ from a set of clauses $S$ if for every
  non-variable term $s$ occurring in $S$, if $t$ and $s$ are unifiable, then
   $t = s$.
   An instantiation scheme $\instgamma$ is {\em admissible} if:
\label{inst_assump}
 \begin{enumerate}
 \item{It is is monotonic, i.e. $S \subseteq S'
     \Rightarrow \instgamma(S) \subseteq
     \instgamma(S')$.\label{inst1}}
 \item{If $S$ is a set of clauses and $t,s$ are two terms \disc\ from
     $S$ then $\instgamma(S \cup \{ t \iseq s \}) = \instgamma(S) \cup
     \{ t \iseq s \}$.\label{inst2}}
 \end{enumerate}

\end{definition}

 The first requirement is fairly intuitive, and is fulfilled by every
 instantiation procedure of our knowledge. The second one states that
 adding equalities between particular terms should not influence the
 instantiation scheme.
 \articlereport{}{This requirement is actually quite strong, as
 evidenced by the following example.

 \begin{example}
   Let $S = \set{p(a,x) , \neg p(b,c)}$. Since $p(a,x)$ and $p(b,c)$
   are not unifiable, an instantiation scheme may not instantiate
   variable $x$ at all. However, by adding  the unit clause $a
   \iseq b$ to this set, the instantiation scheme should instantiate
   $x$ with constant $c$.
 \end{example}
}
 Generic instantiation schemes such as those in
 \cite{LP92,PZ00,GanzKor:InstEq:2004} do not satisfy the second
 requirement. However, it is fulfilled by the one of \cite{EP09}.
% Furthermore, any instantiation scheme satisfying Condition \ref{inst1} above
% can actually be adapted to ensure
% admissibility: it suffices to systematically replace every independent term by a fresh variable
% (in the input clauses and in the instances).

 From now on, we assume that $\instgamma$ denotes an admissible instantiation scheme.
 We show how to extend $\instgamma$ to
 sets of $\dZ$-closed $\dZ$-clauses.
 Such $\dZ$-clauses
 are obtained as the output of the scheme devised in the previous
 section.

 \begin{definition}
   A set of clauses $S = \{ \myabscl{\Lambda_i}{C_i} \mid i \in [1..n]
   \}$ where $\Lambda_i$ is a sequence of ground arithmetic atoms and
   $C_i$ is a clause is \emph{$\instgamma$-compatible}
   if $S' = \{ C_1,\ldots,C_n \} \in \aclass$. \articlereport{}{ In this case,
   $\instgamma(S)$ denotes  the set of ground $\dZ$-clauses $\{
   \myabscl{\Lambda_i}{C_i}\instcl{S'}{C_i} \mid i \in [1..n] \}$.}
\end{definition}

\begin{theorem}\label{th_combscheme}
  Let $S = \{ \myabscl{\Lambda_i}{C_i} \mid i \in [1..n] \}$ denote a
  $\instgamma$-compatible set of $\dZ$-clauses\articlereport{.}{,
    where $\Lambda_i$ is a sequence of ground arithmetic atoms and
    $C_i$ is a clause.}  Let $\chi$ denote a constant not occurring in
  the arithmetic part of the clauses in $S$ or in the scope of a
  function of range $\dZ$ in $S$, and consider a set $G$ of ground
  integer terms such that $\chi$ occurs in no term in $G$.

  Then $S \cup \bigcup_{t \in G} \set{\absclempty{\chi \isless t}}$ is
  $\dZ$-satisfiable if and only if $\instgamma(S)$ is
  $\dZ$-satisfiable.
\end{theorem}
\articlereport{}{
\begin{proof}
   We denote by $\intt$ the set
  of ground integer terms in $S$ that do not contain $\chi$, and by
  $\itt{S}$ the set of integer terms $t$ such that $t$ occurs in $S$
  as an argument of a function whose range is distinct from $\dZ$.  By
  construction, $\itt{S} \subseteq \intt$.

  If $S$ is $\dZ$-satisfiable then it is clear that $\gamma(S)$
is
  $\dZ$-satisfiable.  Now, assume that $\instgamma(S)$ admits a
  $\dZ$-model, which we denote by $I$.  Let $S'$
  (resp. $S_{\instgamma}'$) be the set of clauses $C$ such that $S$
  (resp. $\instgamma(S)$) contains a clause of the form
  $\myabscl{\Lambda}{C}$, where $I \models \Lambda$.  By Condition
  \ref{inst1} on instantiation scheme $\instgamma$, $\instgamma(S')
  \subseteq \instgamma(\{ C_1,\ldots,C_n\})$, hence $\instgamma(S')
  \subseteq S_{\instgamma}'$.  We define the set of equations
  \[ E\ =\ \{ t \iseq s \mid t,s \in \itt{S}, I(t) = I(s) \}.\]
   Since the terms in $\itt{S}$ are all ground, every
  term in $\itt{S}$ is \disc\ from $S$ thus by Condition \ref{inst2},
  $\instgamma(S') \cup E = \instgamma(S' \cup E)$. Furthermore, since
  $I$ is a model of $S_{\instgamma}' \cup E$, necessarily,
  $\instgamma(S' \cup E)$ is satisfiable; and since we assumed that
  the instantiation scheme $\instgamma$ is refutationally complete, so
  is $S' \cup E$.  Let $J$ denote a model of $S' \cup E$.  This
  interpretation is not necessarily a $\dZ$-interpretation, and we
  show how to construct a $\dZ$-interpretation $K$  on
  the same domain as $J$ for all sorts other than $\dZ$, such that $K$
  satisfies $S$.

  Given a function $f$ with profile $\asort_1 \times \ldots \times
  \asort_k \rightarrow \asort$,  $f^K$ is defined as follows.  If $f =
  \chi$ then $K(\chi)$ is an integer strictly greater than every
  integer $I(t)$, where $t \in G \cup \itt{S}$. If $\asort = \dZ$
  and $f \not = \chi$ then $f^K = f^I$.  Finally, if $\asort \not =
  \dZ$ then for every tuple $(d_1,\ldots,d_k)$ in the domain of $K$,
  $f^K(d_1,\ldots,d_k) = f^J(d_1',\ldots,d_k')$ where for every $i\in
  [1..k]$:
  \begin{itemize}
  \item{if $\asort_i \not = \dZ$ then $d_i' = d_i$; }
  \item{if $d_i = K(\chi)$ then $d_i' = J(\chi)$;}
  \item{if $d_i = I(t)$ for some term $t \in \itt{S}$ then $d_i' =
      J(t)$ ($t$ is chosen arbitrarily);}
  \item{$d_i'$ is chosen arbitrarily otherwise.}
  \end{itemize}

  By definition of $K$, $K \models \bigwedge_{t \in G} \neg (\chi
  \isless t)$; furthermore, the interpretations of the integer terms
  in $\intt$, which do
  not contain $\chi$, coincide on $K$ and $I$.  We prove that $K
  \models S$.

  Let $i \in [1..n]$.  If $I \not \models \Lambda_i$ then $K \not
  \models \Lambda_i$ (since $\Lambda_i$ only contains integer terms in
  $\intt$ and $I$ and $K$ agree on such terms), thus $K \models
  \myabscl{\Lambda_i}{C_i}$. We now assume that $I,K \models
  \Lambda_i$, so that $C_i \in S'$.  Let $\sigma$ be a ground
  substitution of domain $\var(C_i)$, we show that $K \models
  C_i\sigma$.

  By definition of $J$, $J \models C_i\sigma$, and $C_i\sigma$
  contains no arithmetic atom. Thus it suffices to prove that for
  every noninteger term $t$ occurring in $C_i\sigma$, we have $K(t) =
  J(t)$.  The proof is by induction on $t$.  Let $t =
  f(t_1,\ldots,t_k)$, by definition $K(f(t_1,\ldots,t_n)) =
  f^K(K(t_1),\ldots,K(t_n)) = f^J(d_1',\ldots,d_k')$ where:
  \begin{itemize}
  \item{if $t_i$ is not of sort $\dZ$ then $d_i' = K(t_i)$. By
      induction hypothesis, $K(t_i) = J(t_i)$.}
  \item{if $t_i$ is of sort $\dZ$ and $K(t_i) \not = K(\chi)$ then
      $t_i$ must occur in $\itt{S}$ thus $K(t_i) = I(t_i)$.  In this case,
      $d_i' = J(t)$ where $t$ is some term in $\itt{S}$ such that
      $K(t_i) = I(t)$. Since $I \models (t \iseq t_i)$ and $t,t_i \in
      \itt{S}$ we have $(t \iseq t_i) \in E$ hence $J(t) =
      J(t_i)$. Thus $K(t_i) = J(t_i)$.}
  \item{If $K(t_i) = K(\chi)$ then $d_i' = J(\chi)$. Since $S$
      contains no integer variable, every ground \emph{integer} term
      in $f(t_1,\ldots,t_n)$ must already occur in $S$. Thus $t_i$
      must occur in $S$ and by definition of $K(\chi)$ we must have
      $t_i = \chi$, hence $d_i' = J(t_i)$.}
\end{itemize}
Thus, $f^K(K(t_1),\ldots,K(t_n)) = f^J(J(t_1),\ldots,J(t_n))$ and
$K(f(t_1,\ldots,t_n)) = J(f(t_1,\ldots,t_n))$.  This implies that
$J(t) = K(t)$ for every noninteger term $t$ occurring in
 $C_i\sigma$. Since $J \models C_i\sigma$, we also
  have $K \models C_i\sigma$, which proves that $S \cup \bigwedge_{t \in
    G} \neg(\chi \isless t)$ is also satisfiable.
\end{proof}
}

\subsubsection*{Summary.}

To summarize, starting from a set of $\dZ$-clauses $S$:

\begin{enumerate}
\item{The scheme devised in Section \ref{sec_instz} is applied to
    instantiate all integer variables occurring in $S$. We obtain a
    $\dZ$-closed set of $\dZ$-clauses $S'$.}
\item $S'$ is processed to get rid of all clauses containing
  arithmetic atoms of the form $\speconst \isless t$, and to get rid
  of all atoms of the form $t\isless
  \speconst$ in the remaining clauses. We obtain a set of
  $\dZ$-clauses $S''$.
\item{Then we apply an admissible
    instantiation scheme (e.g.,
    \cite{EP09}) $\instgamma$ on the clausal part of the $\dZ$-clauses
    in $S''$ to instantiate all remaining variables. We obtain a set
    of closed $\dZ$-clauses $S_g$.}
\item{Finally we feed an SMT-solver (that handles linear arithmetic) with $S_g$. }
\end{enumerate}

The previous results ensure that $S$ and $S_g$ are equisatisfiable,
provided $S''$ is compatible with $\instgamma$. This means that the
procedure can be applied to test the satisfiability of an SMT problem
on the combination of linear arithmetic with, e.g., \emph{any} of the
theories that the scheme of \cite{EP09} is capable of handling, which
include the theories of arrays, records, or lists.  Note that an
efficient implementation of this scheme would not instantiate
variables by $\speconst$ in clauses or literals that are afterwards
deleted, but would directly apply the simplification.

Note also that simple optimizations can further be applied to reduce
the size of the instantiation set. For example, given a set of clauses
$S$, there is no need to keep in the instantiation set $B_S$ two
distinct terms $t$ and $s$ such that $S \modelz t\iseq s$. Thus, it is
useless to store in $B_S$ a constant $a$ and a term $\prc(\su(a))$; if
$S$ contains a unit clause $t\iseq a$, there is no need for $B_S$ to
contain both $t$ and $a$.
Another rather obvious improvement is to use several distinct sorts interpreted as
integers.
Then the arithmetic variables need only to be instantiated by terms of the same sort.
Our results extend straightforwardly to such
settings, but we chose not to directly include these optimizations in
our proofs for the sake of readability.

\section{Applications}

We now show two applications of our technique to solve satisfiability
problems involving integers.

\subsubsection*{Arrays with integer indices.}

The theory of \emph{arrays with integer indices} is axiomatized by the
following set of clauses, denoted by $\thearrz$:
\[\begin{array}{rccclr}
  & \|\rightarrow &\select(\store(x,z,v),z) &\iseq& v & (a_1)\\
  w\isless \prc(z)& \|\rightarrow& \select(\store(x,z,v),w)& \iseq& \select(x,w) & (a_2)\\
  \su(z)\isless w& \|\rightarrow& \select(\store(x,z,v),w)& \iseq& \select(x,w) & (a_3)
\end{array}\]

Instead of clauses $(a_2)$ and $(a_3)$, the standard axiomatization of
the theory of arrays contains $\abscl{w \niseq
  z}{}{\select(\store(x,z,v),w) \iseq \select(x,w)}$. In order to be
able to apply our scheme, we replaced  atom $w \niseq z$ by the
disjunction $w \isless
\prc(z) \vee \su(z) \isless w$, which is equivalent in $\dZ$.\articlereport{}{The standard axiomatization of the theory of arrays is saturated for the superposition calculus (see, e.g., \cite{ARR03}), and a similar result holds for the new axiomatization:

\begin{proposition}\label{prop_thearrz}
  $\thearrz$ is saturated in $\infsys$.
\end{proposition}
\articlereport{}{
\begin{proof}
  Any inference between the axioms generates a clause that can be
  deleted by the \textbf{tautology deletion} rule.
\end{proof}
}
}

We consider SMT problems on arrays with integer indices of a
particular kind:
\begin{definition}
  An \emph{$\thearrz$-inequality problem} is a set of
  $\dZ$-clauses  $\thearrz \cup S_0$ where:
  \begin{itemize}
  \item the only variables occurring in $S_0$ are integer variables,
  \item all non-ground arithmetic atoms occurring in  $S_0$ that are
    not abstraction literals are of the form
    $x\isless t$ or $t\isless x$, where $t$ is either a variable or a
    ground term,
  \item every variable occurring in a term in
    $C \in S_0$ whose head symbol is $\store$ must occur in a
    grounding abstraction literal in $C$.
  \end{itemize}
\end{definition}

Intuitively, these conditions impose that in the corresponding set of
clauses without any integer term abstracted out, the only non-ground
arithmetic atoms are of the form $x\isless t$ or $t\isless x$, and
every term occurring in $S$ whose head symbol is $\store$ must be ground.

\articlereport{}{
\begin{definition}
  Consider a $\dZ$-clause $C = \absclause$.
  \begin{itemize}
  \item $C$ is an \emph{array property clause} if it only contains
    integer variables,  $\Gamma \rightarrow \Delta$ contains no
    occurrence of the $\store$ symbol, and every occurrence of the
    $\select$ symbol admits a constant as a first argument.
  \item $C$ is an \emph{array write clause} if it is of the form
    \[\abscl{\Lambda', u \iseq i}{\Gamma', \store(a,u,e) \iseq
      b}{\Delta},\]
    where $a,e,b$ and all terms in $\Gamma, \Delta$ are flat and ground.
  \end{itemize}
\end{definition}

It is simple to verify that every $\thearrz$-inequality problem can be
reduced by the flattening operation to an equisatisfiable set of
clauses of the form $\thearrz \cup S_p \cup S_w$, where:
\begin{itemize}
\item The clauses in $S_p$ are array property clauses. Intuitively,
  the clauses in these set are used to define properties on the arrays
  under consideration.
\item The clauses in $S_w$ are array write clauses. Intuitively, the
  clauses in these set represent the write operations on the arrays
  under consideration.
\end{itemize}
}

\articlereport{}{

\begin{proposition}\label{prop_array_inf}
  The following results hold:
  \begin{enumerate}
  \item An inference between an array property clause and an array
    write clause has an empty mgu, and it generates an array write
    clause.\label{it_write}
  \item There are no possible inferences between an array property
    clause and an axiom in $\thearrz$.
  \item An inference between an array write clause and an axiom in
    $\thearrz$ generates an array property clause.\label{it_axwrite}
  \end{enumerate}
\end{proposition}

\articlereport{}{
\begin{proof}
  \begin{enumerate}
  \item An array write clause is of the form
    \[\abscl{\Lambda, u \iseq i}{\Gamma, \store(a,u,e) \iseq b}{\Delta},\]
    where every term in $\Gamma, \Delta$ is flat and ground. Since
    there can be no occurrence of $\store$ or of a literal $x\iseq t$
    in the array property clause, its maximal literal must be an
    equation between constants. Hence the mgu of the two unified terms
    is empty and the generated clause is an array write clause.
  \item Since array property clauses do not contain any occurrence of
    $\store$ and no superposition into a variable is permitted, an
    inference between an array property clause and an axiom in
    $\thearrz$ must unify terms with $\select$ as a head symbol. But
    in this case, the unified term in the axiom must be a
    $\select(\store(x,z,v),w)$, where $w$ may be equal to $z$, and
    since the unified term in the array property clause must be of the
    form $\select(a,u,t)$ where $a$ is a constant, these terms cannot
    be unifiable.
  \item The only term on which the superposition rule can apply in an
    array write clause is of the form $\store(a,u,e)$ (recall that
    constant symbols are strictly smaller than complex terms and that no
    equation on integer variables is allowed in the clauses). Since
    $\thearrz$ contains no occurrence of $\store$ at the root level,
    the rule must apply from the array write clause into
    $\thearrz$. Thus it replaces a term $\store(x,z,v)$ occurring in
    the axiom by $b$, which implies that the first argument of
    $\select$ is a constant. Furthermore, $x$ and $v$ are instantiated by
    constants by unification, thus the obtained clause contains no
    variable except for the integer variables $z,w$.
  \end{enumerate}
\end{proof}
}

\begin{proposition}\label{prop_preserv}
  Let $D = \abscl{\Lambda, u\iseq i}{\Gamma, \store(a,u,e) \iseq
    b}{\Delta}$ denote an array write clause generated from $S_p \cup
  S_w$. Then $S_w$ contains an array write clause of the form
  $\abscl{\Lambda', u\iseq i}{\Gamma', \store(a',u,e') \iseq b'}{\Delta'}$.
\end{proposition}

\begin{proof}
  The result is a direct consequence of Proposition
  \ref{prop_array_inf} \myref{it_write}, and is proved by induction on
  the length of the derivation generating $D$.
\end{proof}
}
For every $\thearrz$-inequality problem $S$,
we define the following set of ground terms, which will be used
throughout this section:
\begin{eqnarray*}
  B_S& =& \setof{t \textrm{ ground}}{x \isless t \textrm{ or }
    \select(a,t) \textrm{ occurs in } S}\\
  &\cup & \setof{t'\textrm{ ground}}{\store(a,u, e) \iseq b \textrm{ and } u \iseq t'
    \textrm{ occur in a same clause in } S}\\
  &\cup&
  \setof{\prc(t')\textrm{ ground}}{\store(a,u, e) \iseq b, u \iseq t'
    \textrm{ occur in a same clause in } S}
\end{eqnarray*}

\articlereport{}{
\begin{proposition}
  For every clause $C$ in $S_p \cup S_w$, $C$ is {\preconstrained} and
  $C\spsub B_S$.
\end{proposition}

Note that the clauses in $\thearrz$ are not {\preconstrained}.

\begin{lemma}
  Every non-redundant clause $C$ generated from $\thearrz \cup S_p
  \cup S_w$ other than the clauses in $\thearrz$ is
  {\preconstrained} and such that $C \spsub B_S$.
\end{lemma}

\begin{proof}
  The property holds for the clauses in $S_p \cup S_w$.  We prove the
  result by induction on the length of the derivation, and prove at
  the same time that $C$ is either an array property clause or an
  array write clause; this is trivially true if $C \in S_p \cup
  S_w$. Assume $C$ is generated by a derivation of length 1, i.e., by
  an inference on $D,D'$; these clause are not necessarily
  distinct. We perform a case analysis on the properties $D$ and $D'$
  satisfy:
  \begin{description}
  \item [$D$ and $D'$ are axioms in $\thearrz$.] In this case, $C$ is
    redundant  by Proposition \ref{prop_thearrz}.
  \item [$D$ is an array property clause and $D' \in \thearrz$.] By
    Proposition \ref{prop_array_inf}, this case is not possible.
  \item [$D$ is an array write clause and $D' \in \thearrz$.]  Then
    $D$ is a clause of the form $\abscl{\Lambda, u \iseq
      i}{\Gamma}{\store(a,u,e) \iseq b, \Delta}$, where a term
    $\store(a', i, e')$ occurs in $S_w$ by Proposition
    \ref{prop_preserv}. By Proposition \ref{prop_array_inf}
    \myref{it_axwrite}, $C$ is an array property clause. Assume $D' = (a_2)$,
    the other cases are similar. Then \[C = \abscl{w
      \isless \prc(u), u \iseq i, \Lambda}{}{\select(b,w) \iseq
      \select(a,w)},\] $C$ is therefore an array property clause such that
    $C\spsub B_S$, and it is {\preconstrained}.
  \item [$D$ and $D'$ are in $S_p\cup S_w$.] In this case, $C\spsub
    B_S$ by Lemma \ref{lm_spsub}, $C$ is {\preconstrained} by Lemma
    \ref{lm_pres_prec}, and it is either an array property clause, or
    an array write clause by Proposition \ref{prop_array_inf}.
  \end{description}
\end{proof}

}

\articlereport{The following lemma is a consequence of Theorem \ref{th_inst}.}{ By Theorem \ref{th_inst}, if we consider the set
  $B_S'$ obtained from $B_S$ by adding a constant $\speconst$ not
  occurring in $S$, then $S$ and $\addsetabs{S}{B_S'}\cup \bigcup_{t
    \in B_S}\set{\abscl{\speconst\isless t}{}{}}$ are equisatisfiable.
  We restate this result using substitutions instead of abstraction
  atoms: }
\begin{lemma}
\label{lem:omega}
  Let $B_S' = B_S \cup \set{\speconst}$, let $V$ denote the set of inequality
  variables occurring in clauses in $S$, and let $\Omega$ denote the
  set of all substitutions of domain $V$ and codomain $B_S'$. Then
  $\thearrz \cup S_0$ and $(\thearrz \cup S_0)\Omega$ are
  equisatisfiable.
\end{lemma}

Since we  assumed all integer variables in $S$ are either abstraction
variables or inequality variables (by otherwise adding $x\isless x$ to
the necessary clauses), we conclude that the clauses in $S_0\Omega$ are
all ground, and the clauses in $\thearrz\Omega$ are of the form:
\[\begin{array}{rcrcl}
  &\|&\select(\store(x,z,v),z) &\iseq& v\\
  s\isless \prc(z) &\|& \select(\store(x,z,v),s)& \iseq& \select(x,s)\\
  \su(z)\isless s& \|& \select(\store(x,z,v),s)& \iseq& \select(x,s),
\end{array}\]
where $s$ is a ground term. This set of terms can be instantiated
using the scheme of \cite{EP09}. Thus, if $\Omega'$ denotes the set of
substitutions constructed by the instantiation scheme, by Theorem
\ref{th_combscheme}, the sets $S$ and $S\Omega\Omega'$ are
equisatisfiable. The latter is ground, and its satisfiability can be
tested by any SMT solver capable of handling linear arithmetic and
congruence closure.

We would like to emphasize that similar theories can be handled in the same way, for instance the theory of lists, records, etc. Furthermore, other axioms can be added in the theory of arrays to
express additional properties, such as sortedness (see \cite{EP09} for details).

\subsubsection*{An example.}

Consider the following sets:
\begin{eqnarray*}
  E & =& \setof{l_i \isless x_i \isless u_i\ \|\ \rightarrow \select(a, x_i)
    \iseq e_i}{i = 1, \ldots, n},\\
  F & = & \setof{u_i \isless \prc(l_i)\ \|\ \rightarrow}{i = 1, \ldots, n},\\
  G & = &  \setof{u_i \isless \prc(l_{i+1})\ \|\ \rightarrow}{i = 1, \ldots, n-1},
\end{eqnarray*}
where the $u_i$'s and $l_j$'s are constants. The $\dZ$-clauses in $E$ state
that array $a$ is constant between bounds $l_i$ and $u_i$, for $i = 1,
\ldots, n$; the $\dZ$-clauses in $F$ state that each interval has at least
$1$ element, and the $\dZ$-clauses in $G$ state that all the intervals have
a nonempty intersection. Thus, the union of these sets entails that
$a$ is constant between bounds $l_1$ and $u_n$. Let $b$ denote the
array obtained from $a$ by writing element $e_1$ at position
$u_{n+1}$. If $u_{n+1} = \su(u_n)$, then $b$ is constant between bounds
$l_1$ and $\su(u_n)$. Let
\begin{eqnarray*}
  H & = & \set{x\iseq u_{n+1}\|\ \rightarrow b \iseq
    \store(a,x,e_1),\ \|\ \rightarrow u' \iseq
    \su(u_n)} \\
  & \cup & \set{k\isless \prc(l_1)\ \|\ \rightarrow\ , u_n\isless
    \prc(k)\ \|\ \rightarrow\ }\\
  & \cup & \set{\|\ \select(b,k)
    \iseq e_1\rightarrow} \textrm{ and}\\
  S_0& =& E \cup F \cup G \cup H,
\end{eqnarray*}
then $\thearrz \cup S_0$ is unsatisfiable. By applying the definition
of $B_S$ from the previous section, we obtain $B_S = \set{u_1, \ldots,
  u_n, u_{n+1}, \prc(u_{n+1}), k}$. In the first step, all variables in $E$ are
instantiated with the elements of $B_S' = B_S \cup \set{\speconst}$,
yielding\footnote{in an actual implementation,
  the variables in $E$ would not be instantiated  with $\speconst$.}
a ground set $E'$. The
inequality variables in the axioms of $\thearrz$ are also instantiated
with the elements of $B_S'$, yielding a set of clauses $A$. Then, in the
second step, the clauses in $A$ are instantiated using the term
$\store(a, u_{n+1}, e_1)$, and we obtain a set $A'$ containing
clauses of the form
\[\begin{array}{rcrcl}
 x\iseq u_{n+1}& \| & \select(\store(a, x, e_1), x)
  &\iseq& e_1,\\
  x\iseq u_{n+1},s \isless \prc(x) & \| & \select(\store(a, x, e_1), s)
  &\iseq& \select(a, s),\\
  x\iseq u_{n+1},\su(x) \isless s & \| & \select(\store(a, x, e_1), s)
  &\iseq& \select(a, s),\\
\end{array}\]
where $s \in B_S'$. Then an SMT solver is invoked on the ground set of
clauses $A' \cup E' \cup F\cup G \cup H$. The size of this set is to
be compared with the one obtained by the procedure of
\cite{BradleyBook},  clauses are instantiated using an index set
\articlereport{$\mathcal{I} = \setof{l_i, u_i}{i = 1, \ldots, n} \cup
  \set{u_{n+1}, \prc(u_{n+1}), \su(u_{n+1}), \su(u_n), k}.$}{\begin{eqnarray*}
  \mathcal{I} & = & \setof{l_i, u_i}{i = 1, \ldots, n} \cup
  \set{u_{n+1}, \prc(u_{n+1}), \su(u_{n+1}), \su(u_n), k}.
\end{eqnarray*}}

There are twice as many terms in this instantiation set.
It is simple to check that our procedure
always generates less instances than the one of \cite{BradleyBook}.
In fact, there are cases in which our method is exponentially
better.
\articlereport{For example, }
{The simplest example is the following: }
for $i = 1, \ldots, n$,
let $A_i$ denote the atom $\select(a,x_i) \iseq c_i$, and let $S =
\thearrz \cup S_0$, where
\articlereport{$S_0 = \{\abscl{i \isless
  x_1, \ldots, i \isless x_n, j \isless y}{A_1, \ldots, A_n}{\select(a,y) \iseq
  e} \}.$}{
\begin{eqnarray*}
S_0 &=& \{\abscl{a \leq
  x_1, \ldots, a \leq x_n, b \leq y}{A_1, \ldots, A_n}{\select(t,y) \iseq
  b} \}.
\end{eqnarray*}}
With this set,  our instantiation scheme generates only a
\emph{unique} clause, whereas the one in \cite{BradleyBook}
instantiates every $x_i$ with $i$ and $j$, yielding $2^n$ clauses.

\subsubsection*{Stratified classes.}

To show the wide range of applicability of our results, we provide
another example of a domain where they can be applied.  The results in
this section concern decidable
subclasses of first-order logic with sorts, which are investigated in
\cite{Abadi2010153}.
We briefly review some definitions.
\begin{definition}
A set of function symbols
$\Sigma$ is \emph{stratified} if there exists a function $\lev$
mapping every sort $\asort$ to a natural number such that for every
function symbol $f \in \Sigma$ of profile $\asort_1 \times \ldots
\asort_n \rightarrow \asort$ and for every $i \in [1..n]$,
$\lev(\asort) > \lev(\asort_i)$.
We denote by $T_\Sigma$ (resp. $T_\Sigma^\asort$) the set of ground
terms built on the set of function symbols $\Sigma$ (resp. the set of
terms of sort $\asort$ built on $\Sigma$).
\end{definition}

\begin{proposition}
\label{prop:finitestrat}
Let $\Sigma$ be a finite stratified set of function symbols.
Then the set $T_\Sigma$ is finite.
\end{proposition}

\articlereport{}{
\begin{proof}
  We show, by induction on the terms, that the depth of a term in
  $T_\Sigma^\asort$  is bounded by $\lev(\asort)$. This
  obviously implies that the number of  terms in $T_\Sigma$ is finite.

  Let $t$ be a ground term of depth $d$. $t$ is of the form
  $f(t_1,\ldots,t_n)$ where $f$ is a function symbol of profile
  $\asort_1 \times \ldots \times \asort_n \rightarrow \asort$ and
  $t_1,\ldots,t_n$ are terms of sorts $\asort_1,\ldots,\asort_n$
  respectively.  By the induction hypothesis, the depth of
  $t_1,\ldots,t_n$ is bounded by
  $\lev(\asort_1),\ldots,\lev(\asort_n)$ respectively. Thus the depth
  of $t$ is bounded by $1+\max_{i \in [1..n]}(\lev(\asort_i))$. Since
  the signature is stratified, we have $\forall i \in [1..n],
  \lev(\asort) > \lev(\asort_i)$ thus $\lev(\asort) \geq 1+\max_{i \in
    [1..n]}(\lev(\asort_i)) \geq d$.
\end{proof}
} A set of clauses is in $\stzero$ if its signature is stratified. In
particular, any formula in the Bernays-Sch\"onfinkel class is in
$\stzero$.  By Proposition \ref{prop:finitestrat}, $\stzero$ admits a
trivial instantiation scheme: it suffices to replace each variable by
every ground term of the same sort, defined on the set of function
symbols occurring in $\stzero$\footnote{possibly enriched with some
  constant symbols in order to ensure that each sort is
  nonempty.}. This instantiation scheme is obviously admissible (see
Definition \ref{def_instadmin}).

This instantiation scheme can be applied also to the class $St_2$
defined in \cite{Abadi2010153} as an extension of the class $\stzero$
with atoms of the form $t \in \im{f}$, where $f$ is a function symbol
of profile $\asort_1 \times \ldots \times \asort_n \rightarrow
\asort$, meaning that $t$ is in the image of the function $f$.  From a
semantic point of view, the atom $t \in \im{f}$ is a shorthand for
$\exists x_1,\ldots,x_n. t \iseq f(x_1,\ldots,x_n)$.  To ensure
decidability, for every atom of the form $t \in \im{f}$ and for every
function symbol $g$ of the same range as $f$, the following properties
have to be satisfied: \articlereport{$(i)$ $g$ must have the same
    profile as $f$; $(ii)$ the formula $f(x_1,\ldots,x_n) \iseq
    g(y_1,\ldots,y_n) \Rightarrow \bigwedge_{i=1}^n x_i \iseq y_i$,
    where $n$ denotes the arity of $f$ and $g$, must hold in every
    model of the considered formula.}{
\begin{enumerate}
\item{$g$ must have the same profile as $f$.}
\item{The formula $f(x_1,\ldots,x_n) \iseq g(y_1,\ldots,y_n) \Rightarrow
\bigwedge_{i=1}^n x_i \iseq y_i$, where $n$ denotes the arity of $f$ and $g$, must hold in every model of the considered formula.}
\end{enumerate}
}
In \cite{Abadi2010153} it is shown that every satisfiable set in
$\sttwo$ admits a finite model, hence, $\sttwo$ is decidable.
\articlereport{It turns out that}{We show that} any formula in $\sttwo$ can be reduced to a clause set in
$\stzero$, thus reducing  satisfiability problems in $\sttwo$ to
satisfiability problems in $\stzero$.
\articlereport{}{We begin by showing that if $t$ is a complex term in an
atom $t \in \im{f}$, then under certain conditions which will be
satisfied for the elements in $\sttwo$, the atom can  safely be
replaced by an equality atom.

\begin{proposition}
\label{prop:im}
Consider an atom $g(t_1,\ldots,t_n) \in \im{f}$, and assume that $g$
has the same profile as $f$. Consider also an interpretation $I$  such
that $I \models f(x_1,\ldots,x_n) \iseq g(y_1,\ldots,y_n) \Rightarrow
\bigwedge_{i=1}^n x_i \iseq y_i$.
Then $I \models g(t_1,\ldots,t_n) \in \im{f}$ if and only if $I
\models g(t_1,\ldots,t_n) \iseq f(t_1,\ldots,t_n)$.
\end{proposition}

\articlereport{}{
\begin{proof}
  First note that $f(t_1,\ldots,t_n)$ is a well-formed term, since $f$
  and $g$ have the same profile by hypothesis.  Furthermore, it is
  obvious that $g(t_1,\ldots,t_n) \iseq f(t_1,\ldots,t_n) \models
  g(t_1,\ldots,t_n) \in \im{f}$.  Now, assume that $I \models
  g(t_1,\ldots,t_n) \in \im{f}$. Then there exists an extension $I'$
  of $I$ to $x_1,\ldots,x_n$ such that $I' \models g(t_1,\ldots,t_n)
  \iseq f(x_1,\ldots,x_n)$.  But then since $I \models
  f(x_1,\ldots,x_n) \iseq g(y_1,\ldots,y_n) \Rightarrow
  \bigwedge_{i=1}^n x_i \iseq y_i$ we deduce that $I' \models t_i
  \iseq x_i$, for all $i \in [1..n]$.  Thus $I' \models
  g(t_1,\ldots,t_n) \iseq f(t_1,\ldots,t_n)$, and $I \models
  g(t_1,\ldots,t_n) \iseq f(t_1,\ldots,t_n)$ since $I$ and $I'$ coincide on
  the terms not containing $x_1,\ldots,x_n$.
\end{proof}
}

In order to get rid of atoms of the form $t\in \im{f}$, we prove that
in the case where $t$ is a variable, we may assume $t$ is interpreted
as a ground term in $T_\Sigma$.

\begin{lemma}
\label{lem:modst}
Let $S$ denote a clause set in $\sttwo$ built on a stratified set of
symbols $\Sigma$ such that $T_\Sigma^\asort$ is nonempty for every
$\asort \in \sorts$.  Let $S$ denote a clause set in $\sttwo$.  If an
interpretation $I$ is a model of $S$, then the restriction of $I$ to
the domains $I(T_\Sigma^\asort)$ is also a model of $S$.
\end{lemma}

\articlereport{}{
\begin{proof}
  Let $I$ denote a model of $S$, and let $J$ be the restriction of $I$
  to the domains $I(T_\Sigma^\asort)$, for every sort $\asort \in
  \sorts$. It is clear that $J$ is an interpretation and, by
  definition, for all $f \in \Sigma$, $f^J$ is a total function. Still
  by definition, $I$ and $J$ coincide on every term in $T_\Sigma$.
  Let $\sigma$ denote a grounding substitution and let $A$ denote an
  atom built on the symbols in $\Sigma$.  If $A$ is of the form $t
  \iseq s$ then we have $I(t\sigma) = J(t\sigma)$ and $I(s\sigma) =
  J(s\sigma)$ by construction, thus, $I \models (t \iseq s)\sigma$ if
  and only if $J \models (t \iseq s)\sigma$. Thus ground equational
  atoms have the same truth values in $I$ and in $J$. This implies
  that the purely equational clauses have the same truth values in $I$
  and in $J$, and in particular, we deduce that $J \models
  f(x_1,\ldots,x_n) \iseq g(y_1,\ldots,y_n) \Rightarrow
  \bigwedge_{i=1}^n x_i \iseq y_i$, for every function symbol $f$
  occurring in a term $\im{f}$ and for every function symbol $g$ with
  the same range as $f$.

  If $A$ is of the form $t \in \im{f}$, then by definition $t\sigma$
  is of the form $g(t_1,\ldots,t_n)$ for some function symbol with the
  same range as $f$.  By Proposition \ref{prop:im}, $t\sigma \in
  \im{f}$ has the same truth value in $I$ and in $J$ as $t\sigma \iseq
  f(t_1,\ldots,t_n)$.  Since $t\sigma \iseq f(t_1,\ldots,t_n)$ is a
  ground equational atom, it has the same truth value in $I$ and in
  $J$. Thus $I \models t\sigma \in \im{f}$ if and only if $J \models
  t\sigma \in \im{f}$. We deduce that for all grounding substitutions
  $\sigma$, $A\sigma$ has the same truth value in $I$ and in $J$, and
  since $I\models S$, we conclude that $J\models S$.
\end{proof}
}

Proposition \ref{prop:im} and Lemma \ref{lem:modst} permit to
reduce a satisfiability problem in $\sttwo$ to a satisfiability
problem in $\stzero$, by getting rid of all occurrences of atoms of
the form $t \in \im{f}$.}
{This is done by getting rid of all occurrences of atoms of
the form $t \in \im{f}$.}
  One such transformation is obvious: by
definition, every occurrence of the form $t \not\in \im{f}$ can be
replaced by $t \niseq f(x_1, \ldots, x_n)$, where the $x_i$ are fresh
variables. We now focus on the other occurrences of the atoms.
\begin{definition}\label{def_strat_trans}
  Let $S$ denote a set of clauses. We denote by $S'$ the set of clauses
  obtained from $S$ by applying the following transformation rule
  (using a ``don't care'' nondeterministic strategy):
\vspace{-0.3cm}
\[
\Gamma\! \rightarrow\! \Delta, x \in \im{f} \leadsto \{ x \iseq
g(x_1,\ldots,x_n), \Gamma \rightarrow \Delta, g(x_1,\ldots,x_n) \in
\im{f} \mid \mbox{$g \in \Sigma_f$}\}\vspace{-0.3cm}\]
 where $x$ is a variable, $f$
is of arity $n$,
$\Sigma_f$ denotes the set of function symbols with the same profile
as $f$ and $x_1,\ldots,x_n$ are fresh variables that are pairwise
distinct.
We denote by $\tr{S}$ the set of clauses obtained from $S'$ by
applying the following transformation rule:
$g(x_1,\ldots,x_n) \in \im{f} \leadsto g(x_1,\ldots,x_n) \iseq f(x_1,\ldots,x_n)$.
\end{definition}
The first rule gets rids of atoms of the form $x\in \im{f}$ by
replacing them with atoms of the form $t\in \im{f}$ where $t$ is a
complex term, and the second rule gets rid of these atoms.
\articlereport{}{
It is
obvious that these rules terminate: the first one decreases the number
of atoms of the form $x \in \im{f}$ where $x$ is a variable, and the
second one decreases the number of occurrences of $\im{f}$. Obviously,
the normal forms cannot contain atoms of the form $t \in \im{f}$ thus they
must be in $\stzero$.
}
\articlereport{The rules terminate and preserve satisfiability (see the technical report at \url{http://membres-lig.imag.fr/peltier/inst_la.pdf} for details).}{
The rules preserve satisfiability: Proposition \ref{prop:im} ensures
that the second rule preserves equivalence, and Lemma \ref{lem:modst}
ensures that the first one preserves satisfiability, since it shows
that we can restrict ourselves to models in which the
formula $$\forall x \exists x_1,\ldots,x_n. \bigvee_{g \in \Sigma_f} x
\iseq g(x_1,\ldots,x_n)$$ holds.  Note that Condition $1$ in the
definition of $\sttwo$ ensures that every function symbol of the same
range as $f$ is actually in $\Sigma_f$.}
 We therefore have the
following result:

\articlereport{}
{
\begin{theorem}
\label{th:sttwo}
Let $S \in \sttwo$, then $\tr{S} \in \stzero$. Furthermore, $S$ is
satisfiable if and only if $\tr{S}$ is
satisfiable.
\end{theorem}

In particular, these results hold when one of the sorts under
consideration is $\dZ$ and $S$ contains $\dZ$-clauses:
\begin{corollary}\label{cor_sttwo}
  If $S\in \sttwo$ is a set of $\dZ$-clauses, then $S$ is
  $\dZ$-satisfiable if and only if $\tr{S}$ is $\dZ$-satisfiable.
\end{corollary}
}

\begin{theorem}
  Consider a set of $\dZ$-clauses $S = \{ \myabscl{\Lambda_i}{C_i}
  \mid i \in [1..n] \}$ in $\sttwo$ such that every
  $\myabscl{\Lambda_i}{C_i}$ is {\preconstrained}, and for every
  occurrence of an atom $t \in \im{f}$, the range of $f$ is not of
  sort $\dZ$. The set $S$ is $\dZ$-satisfiable if and only if
  $\instgamma(\tr{S}\Omega)$ is $\dZ$-satisfiable, where:
  \begin{itemize}
  \item{$\Omega$ is the set of substitutions of domain $\var(S)$ whose
      codomain is a set $B$ such that $\myabscl{\Lambda_i}{C_i}
      \spsub B$ for all $i = 1, \ldots, n$;}
  \item{$\instgamma$ denotes an instantiation scheme for $\stzero$
      satisfying the conditions of page \pageref{inst_assump}
      (e.g. $\instgamma(S) = S\Omega'$ where $\Omega'$ is the set of
      substitutions of domain $\var(S)$ and of codomain $T_\Sigma$).}
\end{itemize}
%Thus, the $\dZ$-satisfiability problem is decidable in $\sttwo$.
\end{theorem}

\articlereport{}{
\begin{proof}
  By Corollary \ref{cor_sttwo}, $S$ and $\tr{S}$ are equisatisfiable
  in $\dZ$, and since the transformation rules of Definition
  \ref{def_strat_trans} do not influence the arithmetic parts of the
  $\dZ$-clauses which do not contain any atom of the form $t \in
  \im{f}$, the resulting clauses are {\preconstrained} and upper
  bounded. Thus, by Theorem \ref{th_inst}, $\tr{S}$ and $\tr{S}\Omega
  \cup \bigcup_{t \in B}\set{\absclempty{\speconst \isless t}}$ are
  equisatisfiable. By applying Theorem \ref{th_combscheme}, we deduce
  that $\tr{S}\Omega\cup \bigcup_{t \in B}\set{\absclempty{\speconst
      \isless t}}$ and $\instgamma(\tr{S}\Omega)$ are equisatisfiable,
  hence the result.
\end{proof}}

Examples of specifications in the classes $\stzero$ and $\sttwo$ are
presented in \cite{Abadi2010153}. Our results allow the
integration of integer constraints into these specifications.

\section{Discussion}

In this paper we presented a way of defining an instantiation scheme
for SMT problems based on a combination of linear arithmetic with
another theory. The scheme consists in getting rid of the integer
variables in the problem, and applying another instantiation procedure
to the resulting problem. Provided the integer variables essentially
occur in inequality constraints, this scheme is complete for the
combination of linear arithmetic with several theories of interest to
the SMT community, but also for the combination of linear arithmetic
with other decidable theories such as the class  $\sttwo$ from
\cite{Abadi2010153}.
The application of this scheme to the theory of arrays with integer
indices shows that it can produce considerably fewer ground instances than
other state-of-the-art procedures, such as that of \cite{BradleyBook}.
The instantiation scheme of \cite{EP09} is
currently being implemented, and will be followed by a comparison with
other tools on concrete examples from
SMT-LIB\footnote{\url{http://www.smt-lib.org/}}.

As far as further research is concerned, we intend to investigate how
to generalize this procedure, in which it is imposed that functions of
range $\dZ$ can only have integer arguments. We intend to determine
how to allow other functions of range $\dZ$ while preserving
completeness. It is shown in \cite{BradleyBook} that considering
arrays with integer elements, for which nested reads can be allowed,
gives rise to undecidable problems, but we expect to define decidable
subclasses, that may generalize those in \cite{dMGe}. Dealing with
more general functions of range $\dZ$ should also allow us to devise a
new decision procedure for the class of arrays with dimension that is
considered in \cite{GNRZ07}. We also intend to generalize our approach
to other combinations of theories that do not necessarily involve
linear arithmetic, by determining conditions that guarantee
\emph{combinations} of instantiation schemes can safely be employed to
eliminate all variables from a formula.
Another interesting line of research would be to avoid a systematic grounding of
integer variables
and to use decision procedures for non-ground systems of arithmetic formulae.
The main difficulty is of course that with our current approach,
instantiating integer variables is required to determine how to instantiate the remaining variables.

%{\small
%\bibliography{../../../BIBLIO/biblio}
%\bibliographystyle{abbrv}
%}

\end{document}